\documentclass[10pt]{article}
\usepackage{a4wide}
\usepackage{cite}
\usepackage{amsmath}
\usepackage{amsfonts}
\usepackage{graphicx}

\newcommand{\dd}{{\rm{d}}}        
\newcommand{\Q}{{\mathcal{Q}}}    
\newcommand{\R}{{\mathcal{R}}}    
\newcommand{\T}{{\mathcal{T}}}    
\newcommand{\HH}{{\mathcal{H}}}   
\newcommand{\scri}{{\mathcal{I}}} 
\newcommand{\disc}{{\mathcal{D}}} 

\def \e {{\rm e}}

\def \boldk {\mbox{\boldmath$k$}}
\def \boldl {\mbox{\boldmath$l$}}
\def \boldm {\mbox{\boldmath$m$}}

\newcommand{\rovno}{\!\!\!\!& = &\!\!\!\!}
\newcommand{\equi}{\!\!\!\!& \equiv &\!\!\!\!}

\begin{document}

\title{Accelerating NUT black holes}

\author{
Ji\v{r}\'{\i} Podolsk\'y
and
Adam Vr\'atn\'y\thanks{
{\tt podolsky@mbox.troja.mff.cuni.cz}
and
{\tt vratny.adam@seznam.cz}
}
\\ \ \\ \ \\
Institute of Theoretical Physics, Charles University, \\
V~Hole\v{s}ovi\v{c}k\'ach 2, 18000 Prague 8, Czech Republic.
}

\maketitle

\begin{abstract}
We present and analyze a class of exact spacetimes which describe accelerating black holes with a NUT parameter. First, by two independent methods we verify that the intricate metric found by Chng, Mann and Stelea in 2006 indeed solves Einstein's vacuum field equations of General Relativity. We explicitly calculate all components of the Weyl tensor and determine its algebraic structure. As it turns out, it is actually of algebraically general type~I with four distinct principal null directions. It explains why this class of solutions has not been (and could not be) found within the large Pleba\'nski--Demia\'nski family of type~D spacetimes.

Then we transform the solution into a much more convenient metric form which explicitly depends on three physical parameters: mass $m$, acceleration $\alpha$, and the NUT parameter~$l$. These parameters can independently be set to zero, recovering thus the well-known spacetimes in standard coordinates, namely the $C$-metric, the Taub--NUT metric, the Schwarzschild metric, and flat Minkowski space in spherical coordinates.

Using this new metric, we investigate main physical and geometrical properties of such accelerating NUT black holes. In particular, we localize and study four Killing horizons (two black-hole plus two acceleration horizons) and carefully investigate the curvature. Employing the scalar invariants we prove that there are no curvature singularities whenever the NUT parameter is nonzero. We identify asymptotically flat regions and relate them to conformal infinities. This leads to a complete understanding of the global structure of the spacetimes: each accelerating NUT black hole is a ``throat'' which connects ``our universe'' with a ``parallel universe''. Moreover, the analytic extension of the boost-rotation metric form reveals that there is a pair of such black holes (with four asymptotically flat regions). They uniformly accelerate in opposite directions due to the action of rotating cosmic strings or struts located along the corresponding two axes. Rotation of these sources is directly related to the NUT parameter. In their vicinity there are pathological regions with closed timelike curves.
\end{abstract}

\vfil\noindent
PACS class:  04.20.Jb, 04.70.Bw, 04.70.Dy



\vfil
\eject

\section{Introduction}
\label{intro}

Exact solutions of Einstein's General Relativity play an important role in understanding strong gravity. Among the first and most fundamental such spacetimes, which were found, investigated and understood, were black holes. They exhibit many key features of the relativistic concept of gravity with surprising applications in modern astrophysics. It is now clear that rotating black holes reside in the hearts of almost all galaxies, and that binary black hole systems in the last stage of their evolution are the strongest sources of gravitational waves in our Universe.

In 1976, Pleba\'nski and Demia\'nski \cite{PlebanskiDemianski:1976} presented a nice form of a complete class of exact spacetimes of algebraic type~D (including a double aligned non-null electromagnetic field and any cosmological constant), first obtained by Debever~\cite{Debever:1971} in 1971. This class involves various black holes, possibly charged, rotating and accelerating. In particular, this large family of solutions contains~the well-known Schwarzschild (1915), Reissner--Nordstr\"{o}m (1916--1918), Schwarzschild--de~Sitter (1918), Kerr (1963), Taub--NUT (1963) or Kerr--Newman (1965) black holes, and also the $C$-metric (1918, 1962) which was physically interpreted by Kinnersley--Walker (1970) as uniformly accelerating pair of black holes.

Unfortunately, these interesting types of black holes --- and their combinations --- had to be obtained from the general Pleba\'nski--Demia\'nski metric by special limiting procedures (degenerate transformations), see Section 21.1.2 of the classic compendium~\cite{Stephanietal:2003} for more details. Moreover, it was traditionally believed that the constant coefficients of the two related Pleba\'nski--Demia\'nski quartic metric functions directly encode the physical parameters of the spacetimes.

In 2003, Hong and Teo \cite{HongTeo:2003,HongTeo:2005} came with a simple but very important idea of employing the coordinate freedom to rewrite the $C$-metric in a new form such that its two quartic (cubic in the uncharged case) metric functions are \emph{factorized to simple roots}. This novel approach enormously simplified the associated calculations and --- more importantly --- the physical analysis of the $C$-metric because the roots themselves localize the axes of symmetry and position of horizons.

Inspired by these works of Hong and Teo, with Jerry Griffiths we applied their novel idea to the complete family of Pleba\'nski--Demia\'nski spacetimes \cite{PlebanskiDemianski:1976}. This ``new look'' enabled us to derive an alternative form of this family of type~D black hole solutions, convenient for physical and geometrical interpretation, see \cite{GriffithsPodolsky:2005,GriffithsPodolsky:2006,PodolskyGriffiths:2006} and Chapter~16 of \cite{GriffithsPodolsky:2009} for summarizing review. This form of the metric reads
\begin{eqnarray}
 \dd s^2 \rovno \frac{1}{\Omega^2}\bigg\{-\frac{{\cal Q}}{\varrho^2}
 \Big[\dd t- \Big(a\sin^2\theta +4l\sin^2\!\frac{\theta}{2}\, \Big)\dd\varphi \Big]^2
   +\frac{\varrho^2}{{\cal Q}}\,\dd r^2  \nonumber\\
 && \hspace{6.4mm} +\frac{\varrho^2}{P}\,\dd\theta^2
 +\frac{P}{\varrho^2}\sin^2\theta \Big[ a\dd t
  -\big(r^2+(a+l)^2\big)\dd\varphi \Big]^2 \bigg\},  \label{newPDMetric}
\end{eqnarray}
 where
${P = 1-a_3\cos\theta-a_4\cos^2\theta}$,
${\cal Q} = (\omega^2k+e^2+g^2) -2mr +\epsilon r^2 -2\alpha n\omega^{-1} r^3
   - (\alpha^2k + {\textstyle\frac{1}{3}} \Lambda)r^4$,
   ${\,\Omega=1-\alpha(l+a\cos\theta)\omega^{-1}\,r }$, ${\,\varrho^2 =r^2+(l+a\cos\theta)^2}$,
and ${a_3, a_4, \epsilon, n,k}$  are uniquely determined constants. The free parameters of the solutions have a direct physical meaning, namely the mass~$m$, electric and magnetic charges~$e$ and~$g$, Kerr-like rotation~$a$, NUT-like parameter~$l$, acceleration~$\alpha$, and the cosmological constant~$\Lambda$. All the particular subclasses of the Pleba\'nski--Demia\'nski black holes can be easily obtained from (\ref{newPDMetric}) by simply setting these physical parameters to zero.

At first sight, it would seem possible to obtain an exact vacuum solution for accelerating black holes with a NUT parameter simply by  keeping ${\alpha, m, l}$ and setting ${a=e=g=\Lambda=0}$. However, in \cite{GriffithsPodolsky:2005} we explicitly demonstrated that in such a special case the constant $\alpha$ is a \emph{redundant parameter} which can be removed by a specific coordinate transformation. In other words, the case ${\alpha, m, l}$ is just the ``static'' black hole with a NUT parameter $l$. Thus we argued convincingly in \cite{GriffithsPodolsky:2005} that the solution which would combine the Taub--NUT metric with the $C$-metric is \emph{not included} in the Pleba\'nski--Demia\'nski family of black holes, despite the fact that a more general solution which describes accelerating \emph{and rotating} black holes with NUT parameter is included in it (indeed, in the metric (\ref{newPDMetric}) it is possible to keep $\alpha$, $a$, $l$, $m$ all nonvanishing). This led us in 2005 to a ``private conjecture'' that the genuine accelerating Taub--NUT metric (without the Kerr-like rotation $a$) need not exist at all.

Quite surprisingly, such a solution was found next year in 2006 by  Chng, Mann and Stelea \cite{ChngMannStelea:2006} by applying a sequence of several mathematical generating techniques. It was presented in the following form\footnote{We have only replaced the acceleration parameter $A$ by $\alpha$, and the mass parameter $m$ by $M$.}
\begin{eqnarray}
\dd \bar{s}^2  \rovno - \frac {(y^2 -1)F(y)}{\alpha^2 (x-y)^2} \frac {C^2 \delta}{\bar{H}(x,y)} \left[\mathrm{d}\bar{t} + \frac{1}{C} \left(\frac {(1 - x^2)F(x)}{\alpha^2 (x-y)^2} + \frac{2 M x}{\alpha} \right)\mathrm{d} \varphi \right]^2
\label{metrikaChngMannStelea}\\
&& + \frac{\bar{H}(x,y)}{\alpha^2 (x-y)^2} \left[ (1-x^2) F(x) \mathrm{d} \varphi^2 + \frac{\mathrm{d}x^2}{(1-x^2)F(x)} + \frac{\mathrm{d}y^2}{(y^2-1)F(y)} \right] , \nonumber
\end{eqnarray}
where
\begin{eqnarray}
F(x) \rovno 1 + 2  \alpha M \, x \,, \label{Fx}\\
F(y) \rovno 1 + 2  \alpha M \, y \,, \label{Fy}\\
\bar{H}(x,y) \rovno \frac{1}{2}+\frac{\delta}{2} \left(\frac{(y^2-1)F(y)}{\alpha^2\,(x-y)^2} \right)^2 , \label{barH}
\end{eqnarray}
see Eq.~(35) in~\cite{ChngMannStelea:2006}. This metric explicitly contains four parameters, namely $M$, $\alpha$, $C$ and $\delta$. The authors of \cite{ChngMannStelea:2006} argued that the parameter $\delta$ is related to the NUT parameter in the limiting case when the acceleration vanishes. And, complementarily, when this parameter is set to zero, the $C$-metric can be obtained. It is thus natural to interpret the metric (\ref{metrikaChngMannStelea})--(\ref{barH}) as an exact spacetime with uniformly accelerating black hole \emph{and} a specific twist described by the NUT parameter. This very interesting suggestion surely deserves a deeper analysis. To our knowledge,  during the last 15 years this has not yet been done, and it is the main purpose of this paper.

First, in Sec.~\ref{sec_comments} we will remove the redundant parameter~$C$, simplifying the original metric of~\cite{ChngMannStelea:2006} to the form in which the twist can be set to zero (leading to the standard $C$-metric). Using it, in subsequent Sec.~\ref{sec_vacuum} we will confirm that the metric (\ref{metrikaChngMannStelea})--(\ref{barH}) is indeed a vacuum solution of Einstein's field equations (we will do this by two independent methods, based on the general results summarized of Appendices~A and~B). In Sec.~\ref{sec_algebrtype} we will calculate the NP scalars $\Psi_A$ in a suitable null frame and determine the algebraic type of the Weyl tensor. Since it will turn out to be \emph{algebraically general} with four distinct principal null directions, it \emph{can not belong} to the class of type~D Pleba\'nski--Demia\'nski spacetimes (\ref{newPDMetric}). Then, in Sec.~\ref{sec_newform} we will present a new metric form of the solution which is much better suited for a geometrical and physical interpretation of this class of black holes. When its three parameters $l$, $\alpha$ and~$m$ are set to zero, standard form of the $C$-metric, the Taub--NUT metric, the Schwarzschild metric and eventually Minkowski space are directly obtained. Specific properties of this family of accelerating NUT black holes are investigated in Sec.~\ref{interpretation}. In particular, we study horizons, curvature singularities, asymptotically flat regions, global structure of these spacetimes, and specific nonregularity of the two axes of symmetry, corresponding to rotating cosmic strings or struts (surrounded by regions with closed timelike curves) which are the physical source of acceleration of the pair of black holes.

\newpage

\section{Removing the degeneracy and initial comments}
\label{sec_comments}

We immediately observe that the original metric (\ref{metrikaChngMannStelea}) does not admit setting ${C=0}$ and ${\delta=0}$. The metric degenerates and its investigation is thus complicated. In fact, the constant $C$ is redundant. To solve these problems, we found convenient to perform a transformation of the time coordinate
\begin{equation}
\tau = 2 \lambda \,( \alpha^2C\,\bar{t} - \varphi ) \,,
\end{equation}
where the new real parameter ${\lambda\ge0}$ is defined as
\begin{eqnarray}
\lambda \equiv \frac{ \sqrt{\delta}}{\alpha^2}\,. \label{lambda}
\end{eqnarray}
Rescaling trivially the metric (\ref{metrikaChngMannStelea}) by a \emph{constant} conformal factor, ${\dd \bar{s}^2 \to \dd s^2\equiv 2\,\dd \bar{s}^2}$, we obtain a better representation of the solution
\begin{eqnarray}
\dd s^2  \rovno - \frac {(y^2 -1)F(y)}{\alpha^2 (x-y)^2 H(x,y)}
\left[\dd \tau + 2 \lambda\,F(x)\frac {1-2xy+y^2}{(x-y)^2} \,\dd \varphi \right]^2  \nonumber\\
&& + \frac{H(x,y)}{\alpha^2 (x-y)^2} \left[ (1-x^2) F(x)\, \dd \varphi^2 + \frac{\dd x^2}{(1-x^2)F(x)}
   + \frac{\dd y^2}{(y^2-1)F(y)} \right] , \label{metricAC}
\end{eqnarray}
where the function ${H\equiv 2 \bar{H}}$ takes the form
\begin{equation}
H(x,y) = 1 + \lambda^2\,\frac{(y^2-1)^2 F^2(y)}{(x-y)^4} \,, \label{H}
\end{equation}
and ${F(x)=1+2\alpha M x\,}$, ${\,\,F(y)=1+2\alpha M y\,}$ are the linear functions (\ref{Fx}) and (\ref{Fy}), respectively. Without loss of generality, we may assume ${\alpha\ge0}$.

It is now possible to set ${\lambda=0}$, in which case ${H=1}$, and the new metric reduces to a diagonal line element
\begin{eqnarray}
\dd s^2  = \frac{1}{\alpha^2 (x-y)^2} \left[ - (y^2 -1)F(y)\, \dd \tau^2 +  (1-x^2) F(x)\, \dd \varphi^2
   + \frac{\dd x^2}{(1-x^2)F(x)} + \frac{\dd y^2}{(y^2-1)F(y)} \right] . \label{Cmetric}
\end{eqnarray}
This is the \emph{usual form of the $C$-metric}, see e.g. Eqs. (14.3), (14.4) in  \cite{GriffithsPodolsky:2009} with the identification ${G(x)\equiv(1-x^2)F(x)}$, ${y \to - y}$ and ${m\equiv M}$. In such a  special case, the metric represents a spacetime with pair of Schwarzschild-like black holes of mass ${M\ge0}$ and uniform acceleration $\alpha$ caused by cosmic strings or struts.

The full metric (\ref{metricAC}) with a \emph{generic} $\lambda$ is clearly a \emph{one-parameter generalization} of this $C$-metric. Additional off-diagonal metric component ${\dd t\,\dd\varphi}$ also occurs, indicating that the parameter $\lambda$ is related to an inherent twist/rotation effect in the spacetime. It will be explicitly demonstrated in Sec.~\ref{sec_newform} that this parameter is directly proportional to the genuine NUT parameter~$l$.

Preliminary physical interpretation of (\ref{metricAC}) can now also be done using similar arguments as those for the $C$-metric, as summarized in Chapter~14 of \cite{GriffithsPodolsky:2009}. In particular, we can comment on the character of coordinate singularities. In order to keep the \emph{correct metric signature} of (\ref{metricAC}) and obtain the usual black-hole interpretation of the spacetime, it is necessary to require ${(1-x^2)F(x)\ge0}$. In view of the roots, this restricts the range of the spatial coordinate to ${x\in[-1,1]}$ and puts the constraint ${0\le2\alpha M<1}$. The coordinate singularities at ${x=\pm 1}$ are the \emph{two poles (axes)}. On the other hand, the admitted zeros of the function ${(y^2-1)F(y)}$ represent the \emph{horizons}, and $F(y)$ can be both positive and negative. More arguments on this will be given in Sec.~\ref{interpretation}, where it will also be demonstrated that the singularity of the metric (\ref{metricAC}) at ${x=y}$ corresponds to asymptotically flat \emph{conformal infinity~${\cal I}$}.

\newpage

\section{Checking the vacuum equations}
\label{sec_vacuum}

Next, it is desirable to verify that the metric (\ref{metricAC}) with (\ref{Fx}), (\ref{Fy}), (\ref{H}) is an exact solution of vacuum Einstein's field equations.

With trivial identification ${\tau \equiv t}$, this metric clearly belongs to the \emph{generic class of stationary axially symmetric metrics}
\begin{equation}
\dd s^2  =  g_{tt}\,\dd t^2 + 2 g_{t\varphi}\,\dd t\,\dd\varphi +g_{\varphi\varphi}\,\dd\varphi^2
   + g_{xx}\,\dd x^2 + g_{yy}\,\dd y^2 \,, \label{axisimetric}
\end{equation}
in which all the functions are independent of the temporal coordinate $t$ and angular coordinate $\varphi$.
Indeed, the explicit metric coefficients of the spacetime (\ref{metricAC}) are
\begin{eqnarray}
g_{tt} \rovno - \frac {(y^2 -1)F(y)}{\alpha^2 (x-y)^2 H(x,y)} \,, \nonumber \\
g_{t \varphi} \rovno - 2 \lambda \, \frac {(y^2 -1)F(y)F(x)(1-2 x y+y^2)}{\alpha^2 (x-y)^4H(x,y)} \,, \nonumber \\
g_{\varphi \varphi} \rovno - 4 \lambda^2 \frac {(y^2 -1)F(y)F^2(x)(1-2 x y+y^2)^2 }{\alpha^2 (x-y)^6 H(x,y)}
  + \frac{H(x,y) (1-x^2) F(x) }{\alpha^2 (x-y)^2} \,, \qquad \label{gab}\\
g_{xx} \rovno  \frac{H(x,y)}{\alpha^2 (x-y)^2 (1-x^2)F(x)} \,, \nonumber \\
g_{yy} \rovno  \frac{H(x,y)}{\alpha^2 (x-y)^2 (y^2-1)F(y)} \,. \nonumber
\end{eqnarray}
Interestingly, the subdeterminant
\begin{eqnarray}
D \equiv g_{tt} \, g_{\varphi \varphi} - g_{t \varphi}^2 < 0 \,, \label{defD}
\end{eqnarray}
turns out to be very simple, namely
\begin{eqnarray}
D = - \frac{(1-x^2)F(x)(y^2-1)F(y)}{\alpha^4 (x-y)^4} \,. \label{D}
\end{eqnarray}

Using the expressions (\ref{axisimetric})--(\ref{D}), we need to evaluate the Riemann and Ricci curvature tensors. Unfortunately, standard computer algebra systems did not provide us the results (even after several hours of calculation on a standard desktop PC) when we attempted to perform a \emph{direct calculation} starting from (\ref{gab}). Therefore, we had to employ a more sophisticated approach. Actually, we developed \emph{two independent methods}.

\subsection{Method A}

It turned out much more convenient first to analytically derive explicit expressions for the Christoffel symbols and subsequently the corresponding components of the curvature tensors of the \emph{generic} stationary axisymmetric metric (\ref{axisimetric}). These results are summarized in Appendix~A.

Moreover, instead of using standard textbook definitions of the Riemann and Ricci tensors, we employed their alternative (and equivalent) versions (\ref{Riemann_formula}), (\ref{Ricci_formula})
The main advantage of this approach is that the second derivatives of the metric are all involved explicitly in the simplest possible way. It is not necessary to differentiate the Christoffel symbols which also contain the inverse metric and thus their first derivatives unnecessarily complicate the evaluation of the curvature.

In the second step, we then substituted the explicit metric functions (\ref{gab}), (\ref{D}) into the general expressions (\ref{CHS2}), (\ref{Riemann}) and (\ref{Ricci}). With a usual PC, such a symbolic-algebra computational process using Maple lasted only around 40 seconds. The result of this computation \emph{confirmed that all the Ricci tensor components} (\ref{Ricci}) \emph{are zero}. The metric (\ref{metricAC}) is thus indeed a vacuum solution in Einstein's gravity theory.

\newpage
\subsection{Method B}

To verify this result (and fasten the computation), we also employed an alternative method based on the ``conformal trick''. Its main idea is that, by multiplying the physical metric (\ref{metricAC}) by a suitable conformal factor $\Omega^2$, the metric components of the related unphysical metric become \emph{polynomial expressions}. Their differentiation and combination, which are necessary to evaluate the curvature tensors, are performed much faster. Specifically, we introduced an unphysical metric $\tilde g_{ab}$ via the conformal relation
\begin{eqnarray}
\tilde g_{ab}  \rovno \Omega^2\, g_{ab} \,,
\label{confmetric}
\end{eqnarray}
where
\begin{equation}
\Omega^2 \equiv \alpha^2 \, (1-x^2)F(x) \, (y^2-1) F(y) \, (x-y)^6 \, \tilde{H}(x,y)\,,
\label{Omegaaxisym}
\end{equation}
and
\begin{equation}
\tilde{H}(x,y) \equiv (x-y)^4H(x,y)  = (x-y)^4 + \lambda ^2 \, (y^2-1)^2 F^2(y)  \,.
\label{tildeH}
\end{equation}
The metric functions $\tilde g_{ab}$ are then only polynomials of $x$ and $y$,
\begin{eqnarray}
\tilde{g}_{tt} \rovno - (1-x^2)F(x) (y^2-1)^2 F^2(y) (x-y)^8 \,, \nonumber \\
\tilde{g}_{t \varphi} \rovno -  2 \lambda \, (1-x^2) F^2(x) (y^2-1)^2 F^2(y) (1-2 x y+y^2) (x-y)^6\,, \nonumber \\
\tilde{g}_{\varphi \varphi} \rovno -4 \lambda^2 \, (1-x^2) F^3(x) (y^2-1)^2 F^2(y) (1-2 x y +y^2)^2 (x-y)^4\,, \label{conformalmetric}\\
&&  +  (1-x^2)^2 F^2 (x) (y^2-1) F(y) \, \tilde{H}^2 (x,y) \,, \nonumber \\
\tilde{g}_{xx} \rovno (y^2-1) F(y) \, \tilde{H}^2 (x,y)  \,, \nonumber \\
\tilde{g}_{yy} \rovno  (1 - x^2)F(x) \, \tilde{H}^2 (x,y) \,. \nonumber
\end{eqnarray}
Using the expressions summarized in Appendix~A, we first computed the Christoffel symbols $\tilde{\Gamma}^a _{\ bc}$ and the Ricci tensor components $\tilde{R}_{ab} $ for this conformal metric $\tilde g_{ab}$  (it also has the stationary axisymmetric form (\ref{axisimetric}), only the tilde symbol is added everywhere). Then we employed the expressions (\ref{Ricci_conformal})--(\ref{deftildeD}) derived in Appendix~B to calculate the Ricci tensor components $R_{ab}$ of the physical metric $g_{ab}$, which is (\ref{gab}). The computer algebra manipulation using Maple again verified that ${R_{ab}=0}$, confirming that the metric is a vacuum solution of Einstein's equations.
In fact, the conformal Method~B is faster than Method~A: the computation took only 15 seconds.

\section{Determining the algebraic type of the spacetime}
\label{sec_algebrtype}

It is now necessary to determine the algebraic type of the spacetime which is given by the \emph{algebraic structure of the Weyl tensor}. The standard procedure is to evaluate all its ten components \cite{GriffithsPodolsky:2009, Stephanietal:2003}
\begin{eqnarray}
\Psi_0 \equi  C_{abcd}\, k^{a} m^{b} k^{c} m^{d}\,, \nonumber\\
\Psi_1 \equi  C_{abcd}\, k^{a} l^{b} k^{c} m^{d}\,, \nonumber\\
\Psi_2 \equi C_{abcd}\, k^{a} m^{b} {\bar m}^{c} l^{d}\,,  \label{defPsi}\\
\Psi_3 \equi  C_{abcd}\, l^{a} k^{b} l^{c} {\bar m}^{d}\,, \nonumber\\
\Psi_4 \equi  C_{abcd}\, l^{a} {\bar m}^{b} l^{c} {\bar m}^{d}\,, \nonumber
\end{eqnarray}
in properly normalized \emph{null tetrad} ${\{\boldk , \boldl , \boldm, \bar{\boldm}\}}$. We adopt the most natural tetrad for the metric (\ref{axisimetric}) in the coordinates ${(t, \varphi, x, y)}$, namely
\begin{eqnarray}
\boldk  \equi  \frac{1}{\sqrt{2}}\,\bigg( \frac{1}{\sqrt{-g_{tt}}}\,\partial_t
    + \frac{1}{\sqrt{g_{yy}}}\,\partial_y \bigg) , \nonumber \\
\boldl  \equi  \frac{1}{\sqrt{2}}\,\bigg( \frac{1}{\sqrt{-g_{tt}}}\,\partial_t
    - \frac{1}{\sqrt{g_{yy}}}\,\partial_y \bigg) , \label{tetrad}\\
\boldm \equi \frac{1}{\sqrt{2}}\,\bigg( \sqrt{\frac{g_{tt}}{D}}\,\partial_\varphi
 +\frac{g_{t \varphi}}{\sqrt{Dg_{tt}}}\,\partial_t - \frac{{\rm i}}{\sqrt{g_{xx}}}\,\partial_x \bigg) , \nonumber
\end{eqnarray}
with $D$ given by (\ref{defD}). All the scalar products vanish, except for
\begin{equation}
\boldk \cdot \boldl  = -1\,,\qquad\qquad \boldm\cdot \bar{\boldm}=1\,.
\label{nulnorm}
\end{equation}

For vacuum solutions, the Ricci tensor and Ricci scalar vanish. The Weyl tensor is thus identical to the Riemann curvature tensor, and in expressions (\ref{defPsi}) we can replace $C_{abcd}$ by $R_{abcd}$. In view of the vanishing components of the null tetrad vectors (\ref{tetrad}) and the vanishing components of the Riemann tensor (\ref{Riemann}) of the metric (\ref{axisimetric}), summarized in Appendix~A, the following formulas for the Weyl scalars can be derived
\begin{eqnarray}
\Psi_0
 \rovno \frac{1}{4} \bigg[\,
 \frac{1}{D\, g_{yy}} \, \bigg( \frac{g_{t \varphi}^2 }{g_{tt}} \, R_{tyty} - 2\, g_{t \varphi}\, R_{ty \varphi y} + g_{tt}\, R_{\varphi y \varphi y} \bigg) - \frac{1}{D}\, R_{t \varphi t \varphi}
 \nonumber \\
 && \quad + \frac{1}{g_{xx}} \, \bigg( \frac{1}{g_{tt}} \, R_{txtx} - \frac{1}{g_{yy}} \, R_{xyxy} \bigg) \bigg] \nonumber \\
  && \quad - \, \frac{{\rm i}}{2}\frac{1}{\sqrt{-D}} \frac{1}{\sqrt{g_{xx} \, g_{yy} }} \bigg( \frac{g_{t \varphi} }{g_{tt}}\, R_{txty} - R_{t \varphi x y} - R_{tx \varphi y} \bigg),
 \nonumber \\[0.25cm]
\Psi_1
 \rovno \frac{1}{2} \bigg[\, \frac{1}{ \sqrt{-D}\,g_{yy}} \bigg(R_{ty \varphi y}-\frac{g_{t \varphi}}{g_{tt}} \, R_{tyty}\bigg)
 - \, \frac{{\rm i}}{g_{tt}\sqrt{g_{xx} \, g_{yy}}}\, R_{txty} \bigg],
 \label{Psi-Axisym} \\[0.25cm]
\Psi_2
 \rovno \frac{1}{4} \bigg[\,
 \frac{1}{D\, g_{yy}} \, \bigg( \frac{g_{t \varphi}^2 }{g_{tt}} \, R_{tyty} - 2\, g_{t \varphi}\, R_{ty \varphi y} + g_{tt}\, R_{\varphi y \varphi y} \bigg) + \frac{1}{D}\, R_{t \varphi t \varphi}
 \nonumber \\
 && \quad + \frac{1}{g_{xx}} \, \bigg( \frac{1}{g_{tt}} \, R_{txtx} + \frac{1}{g_{yy}} \, R_{xyxy} \bigg) \bigg] \nonumber \\
 && \quad - \, \frac{{\rm i}}{2}\frac{1}{\sqrt{-D}} \frac{1}{\sqrt{g_{xx} \, g_{yy} }} \bigg( \frac{g_{t \varphi} }{g_{tt}}\, R_{txty} + R_{t \varphi x y} - R_{tx \varphi y} \bigg),
 \nonumber \\[0.25cm]
\Psi_3 \rovno \Psi_1 \,,
 \nonumber \\[0.00cm]
\Psi_4 \rovno \Psi_0 \,. \nonumber
\end{eqnarray}
Notice that, interestingly, the long expressions for $\Psi_0$ and $\Psi_2$ are very similar. In fact, they only differ in \emph{signs of three terms}.

Now, by substituting the explicit components (\ref{gab}) of the metric and the corresponding Riemann tensor (\ref{Riemann}) into (\ref{Psi-Axisym}), the computer algebra system Maple rendered the following Weyl scalars:
\begin{eqnarray}
\Psi_0 = \Psi_4 \rovno -3\,\alpha^2\lambda \, (1-x^2)F(x) \, (y^2-1)F(y)\, \Xi(x,y) \,,
  \nonumber\\[0.25cm]
\Psi_1 = \Psi_3 \rovno -3\,\alpha^2\lambda \,{\rm i}\,\sqrt{(1-x^2)F(x)}\,\sqrt{(y^2-1)F(y)}\, \Sigma(x,y)\, \Xi(x,y) \,,
  \label{eq:Psi} \\[0.25cm]
\Psi_2 \rovno   \Big[\,\alpha^2 \lambda  \, \Pi(x,y) + {\rm i}\,\alpha^3 M   (x-y)^5  \Big] \, \Xi(x,y) \,,
  \nonumber
\end{eqnarray}
where the functions $\Xi$, $\Sigma$ and $\Pi$ are defined as
\begin{eqnarray}
\Xi(x,y)\rovno \frac{\big(H-4\big)\sqrt{H-1} +{\rm i}\,\big(4-3H\big)}{(x-y)^2 H^3} \,,
   \nonumber\\[0.2cm]
\Sigma(x,y) \rovno xy - 1 - \alpha M x\,(1-3y^2) - \alpha M y\, (1+y^2) \,,
   \label{K} \\[0.25cm]
\Pi(x,y) \rovno 2\,\Sigma^2(x,y) - \big[(1-x^2)F(x)-\alpha M (x-y)^3 \big]\,(y^2-1)F(y) \,,
\nonumber
\end{eqnarray}
with ${H\equiv H(x,y)}$ given by (\ref{H}), and ${F(x), F(y)}$ by (\ref{Fx}), (\ref{Fy}). Surprisingly, the key function $\Xi(x,y)$ which \emph{factorizes all the Weyl scalars} can be written in an explicit and compact form as
\begin{equation}
\Xi = \frac{ {\rm i}\,(x-y)^4}{\big[(x-y)^2-\lambda\,{\rm i}\,(y^2-1)(1 + 2  \alpha M y )  \big]^3} \,. \label{XiEXPLICIT}
\end{equation}

From these curvature scalars, we then computed the \emph{scalar invariants} $I$ and $J$, defined as
\begin{equation}
I \equiv \Psi_0 \Psi_4-4 \Psi_1 \Psi_3 + 3\Psi_2 ^2 \,, \qquad
J \equiv \begin{vmatrix}
\Psi_0 & \Psi_1 & \Psi_2  \\
\Psi_1 & \Psi_2 & \Psi_3  \\
\Psi_2 & \Psi_3 & \Psi_4  \\ 
\end{vmatrix}
\,, \label{IJ}
\end{equation}
and using Maple we verified that the equality ${I^3 = 27 J^2}$ \emph{does not hold}. This means (see \cite{Stephanietal:2003, GriffithsPodolsky:2009}) that the metric (\ref{metricAC}) is \textit{algebraically general}, that is of type~I.

Consequently, \emph{the accelerating NUT metric} (\ref{metricAC}) \emph{can not be included in the Pleba\'nski--Demia\'nski family} because this is of algebraic type~D.

Of course, this conclusion is only valid when ${\lambda \not = 0}$. In the case of vanishing $\lambda$, implying ${H=1}$ and thus ${\Xi={\rm i}/(x-y)^2}$, the only nontrivial Weyl scalar remains ${\Psi_2 = - M\alpha^3(x-y)^3}$. Such spacetime is of algebraic type~D, with double degenerate principal null directions $\boldk$ and $\boldl$. In fact, it is the $C$-metric (\ref{Cmetric}) which belongs to the
Pleba\'nski--Demia\'nski class.

Deeper analysis of the algebraic structure will be presented in Secs.~\ref{subsec:curvature} and~\ref{subsec:singularity}.

\subsection{The principal null directions}
\label{PND}

Actually, it is possible to determine \emph{four principal null directions} (PNDs) of the Weyl tensor, and to prove explicitly that they are \emph{all distinct}.

As usual \cite{Stephanietal:2003, GriffithsPodolsky:2009}, we employ the dependence of the Weyl scalars (\ref{defPsi}) on the choice of the null tetrad, namely their transformation properties under a null rotation which keeps $\boldl $ fixed,
\begin{equation}
\boldk'=\boldk+K\,\bar{\boldm}+\bar K\,\boldm+K\bar K\,\boldl,
 \qquad \boldl'=\boldl, \qquad \boldm'=\boldm+K\,\boldl\,,
\label{nullrotation}
\end{equation}
where $K$ is a complex parameter. The component $\Psi_0$ then transforms to
\begin{equation}
{\Psi_0}'={\Psi_0}+4K{\Psi_1}+6K^2{\Psi_2}+4K^3{\Psi_3}+K^4{\Psi_4} \,.
\label{nullrotationPsi0}
\end{equation}
The condition for $\boldk'$ to be a principal null direction is ${{\Psi_0}'=0}$, which is equivalent
\begin{equation}
 {\Psi_0}+4K{\Psi_1}+6K^2{\Psi_2}+4K^3{\Psi_3}+K^4{\Psi_4}=0\,.
\label{Weylquartic}
\end{equation}
Since this is a \emph{quartic} expression in~$K$, there are \emph{exactly four complex roots} $K_i$ (${i=1,2,3,4}$) to this equation. Each $K_i$ corresponds via (\ref{nullrotation}) to the principal null direction ${\boldk'_i}$.

In the case of the metric (\ref{metricAC}), the Weyl scalars with respect to the null tetrad (\ref{tetrad}) are (\ref{eq:Psi}). Due to the special property ${\Psi_4=\Psi_0}$ and ${\Psi_3=\Psi_1}$, the key algebraic equation (\ref{Weylquartic}) simplifies to
\begin{equation}
\Psi_0 \Big(K^2 +\frac{1}{K^2} \Big) +4\Psi_1 \Big(K +\frac{1}{K} \Big) + 6 \Psi_2 = 0 \,,
\label{kvarticka}
\end{equation}
($K$ must be nonvanishing in (\ref{Weylquartic}) because ${\Psi_0 \neq 0}$). It is convenient to introduce a new parameter
\begin{eqnarray}
\kappa \equiv K+\frac{1}{K}\,,
\label{K z upsilon}
\end{eqnarray}
so that (\ref{kvarticka}) reduces to the quadratic equation in $\kappa$,
\begin{eqnarray}
\Psi_0 \,\kappa^2 +4\Psi_1 \,\kappa + 2 (3\Psi_2-\Psi_0) = 0 \,,
\label{kvadraticka}
\end{eqnarray}
with two solutions
\begin{eqnarray}
\kappa_{1,2} = \frac{- 2\Psi_1 \pm \sqrt{4\Psi_1 ^2 -2\Psi_0(3\Psi_2-\Psi_0)}}{\Psi_0} \,.
\label{kvadrsol}
\end{eqnarray}
Finally, we find the roots $K_i$ by solving (\ref{K z upsilon}), that is the quadratic equation ${K^2-\kappa\, K +1 = 0}$:
\begin{eqnarray}
K_i = \frac{\kappa \pm \sqrt{\kappa ^2 -4}}{2} \,,
\label{kvartrsol}
\end{eqnarray}
where  ${\kappa = \kappa_1}$ and ${\kappa = \kappa_2}$. Indeed, we have thus obtained \emph{four explicit complex roots} $K_i$ corresponding to four distinct PNDs ${\boldk'_i}$, which can be expressed using (\ref{nullrotation}).

\newpage

\section{A new convenient form of the metric}
\label{sec_newform}

The metric (\ref{metrikaChngMannStelea}) can be put in an alternative form which is suitable for its physical interpretation, in particular for determining the meaning of its three free parameters. This is achieved by performing the coordinate transformation
\begin{eqnarray}
x= - \cos \theta \,, \quad y = -\frac{1}{\alpha\, (r-r_{-})} \,, \quad \bar{t} = \frac{ r_{+}-r_{-} }{2 \alpha  l  C} \, t \,. \label{transformation_to_newform}
\end{eqnarray}
We introduce the \emph{NUT parameter} $l$ as
\begin{eqnarray}
l \equiv \lambda\,r_{+} = \frac{ \sqrt{\delta}}{\alpha^2}\,r_{+} \,, \label{l}
\end{eqnarray}
using the definition (\ref{lambda}), and a new real \emph{mass parameter} $m$ via the relation
\begin{eqnarray}
m = \sqrt{M^2 - l^2} \,. \label{m}
\end{eqnarray}
Specific combinations of $m$ and $l$ can conveniently be defined and denoted as
\begin{eqnarray}
r_{+} \equi m + \sqrt{m^2+l^2}\,, \nonumber\\
r_{-} \equi m - \sqrt{m^2+l^2}\,,  \label{defr+r-}
\end{eqnarray}
so that  $r_+$ is \emph{always positive} while $r_-$ is \emph{always negative}. Actually, it will soon be seen that these constants describe the location of two Taub--NUT horizons. From these definitions, important identities immediately follow, namely
\begin{eqnarray}
r_{+} + r_{-}    \rovno  2m\,, \nonumber\\ 
r_{+} - r_{-}    \rovno  2\sqrt{m^2+l^2} = 2M \ge 0 \,, \nonumber\\ 
r_{+}r_{-}       \rovno  - l^2\,,  \label{identity4}\\
r_{+}(r_{+}-r_{-}) \rovno  r_{+}^2+l^2\,. \nonumber 
\end{eqnarray}
The original metric (\ref{metrikaChngMannStelea}) with (\ref{Fx})--(\ref{barH}) then becomes
\begin{eqnarray}
\dd \bar{s}^2  \rovno  \frac{1}{\Omega^2}\,
   \Bigg[ -\frac{(r_{+}-r_{-})^2}{2r_{+}^2}\big(1-\alpha^2 (r - r_{-} )^2 \big) \frac{F(y)}{H(x,y)} \nonumber\\
&& \hspace{31.8mm}\times \bigg(\dd t
       - 2l \Big(\cos\theta - \alpha\,\frac{(r - r_{-})^2 F(x) \sin^2\!\theta }{(r_{+} - r_{-})\,\Omega^2}\,\Big)\dd \varphi \bigg)^2 \label{intermedmetric} \\
&&\hspace{4mm}
   + {\textstyle\frac{1}{2}}(r - r_{-} )^2 H(x,y)\bigg( \frac{\dd r^2}{F(y)(r - r_{-} )^2\big(1-\alpha^2 (r - r_{-} )^2 \big)}
   +  \frac{\dd\theta ^2}{F(x)} +F(x) \sin^2\!\theta \,\dd\varphi^2 \bigg)\Bigg] , \nonumber
\end{eqnarray}
where ${\Omega\equiv 1-\alpha \, (r - r_{-}) \cos\theta }$. Of course, the metric functions $F(x)$, $F(y)$ and ${H(x,y)\equiv 2 \bar{H}}$, given by (\ref{Fx}), (\ref{Fy}) and (\ref{H}), respectively,  must be expressed in terms of the new coordinates~$r$ and~$\theta$. It is useful do relabel them as
\begin{eqnarray}
F(x) \to P(\theta)     \rovno  1-\alpha \,(r_{+} - r_{-}) \cos\theta \,, \nonumber\\
F(y) \to F(r)          \rovno  \frac{r-r_{+}}{r-r_{-}} \,, \label{F->F}\\
H(x,y) \to H(r,\theta) \rovno  1+\frac{l^2}{r_{+}^2} \frac{(r-r_{+})^2}{(r-r_{-})^2}
            \frac{\big[1-\alpha^2 (r - r_{-} )^2 \big]^2 }{\big[1-\alpha \, (r - r_{-}) \cos\theta\big]^4} \,. \nonumber
\end{eqnarray}
Notice that $H$ is always positive. Finally, it is natural to introduce two new functions replacing $F(r)$ and $H(r,\theta)$, namely
\begin{eqnarray}
\Q(r)            \equi  F(r)\, (r - r_{-} )^2\, \big(1-\alpha^2 (r - r_{-} )^2 \big)  \,, \nonumber\\
\R^2(r,\theta)   \equi  \frac{r_{+}}{r_{+}-r_{-}}\, (r-r_{-})^2\, H(r,\theta) \,, \label{newfunct}
\end{eqnarray}
and to perform a trivial \emph{rescaling of the whole metric} by a \emph{constant conformal factor} as
\begin{eqnarray}
\dd s^2\equiv \frac{2\,r_{+}}{r_{+}-r_{-}} \,\dd \bar{s}^2   \,. \label{rescaling}
\end{eqnarray}
Thus, the exact solution found in \cite{ChngMannStelea:2006} simplifies considerably to a \emph{new convenient form of the metric}
\begin{eqnarray}
\dd s^2 \rovno  \frac{1}{\Omega^2}\,
   \Bigg[ -\frac{\Q}{\R^2} \bigg(\dd t
       - 2l \big(\cos\theta - \alpha\,\T\sin^2\!\theta \,\big)\dd \varphi \bigg)^2 \nonumber \\
&&\hspace{22.0mm}
   + \frac{\R^2}{\Q} \,\dd r^2
   + \R^2 \bigg( \frac{\dd\theta ^2}{P} +P \sin^2\!\theta \,\dd\varphi^2 \bigg)\Bigg] , \label{newmetric}
\end{eqnarray}
where
\begin{eqnarray}
\Omega(r,\theta) \rovno  1-\alpha \, (r - r_{-}) \cos\theta      \,,\nonumber\\
P(\theta)        \rovno  1-\alpha \, (r_{+} - r_{-}) \cos\theta  \,,\nonumber\\
\Q(r)            \rovno  \big(r-r_{+}\big)\big(r-r_{-}\big)
                         \big(1-\alpha (r-r_{-})\big) \big(1+\alpha (r-r_{-})\big), \label{newfunctions}\\[4pt]
\T(r,\theta)     \rovno  \frac{(r - r_{-})^2 P}{(r_{+} - r_{-})\,\Omega^2}     \,,\nonumber\\
\R^2(r,\theta)   \rovno  \frac{1}{r_{+}^2+l^2} \bigg(
            r_{+}^2(r-r_{-})^2 + l^2(r-r_{+})^2
            \frac{\big[1-\alpha^2 (r - r_{-} )^2 \big]^2 }{\big[1-\alpha \, (r - r_{-}) \cos\theta\big]^4} \bigg). \nonumber
\end{eqnarray}

This new metric form can be used for investigation of geometric properties of the spacetime and for its physical interpretation. It \emph{explicitly} contains \emph{3 free parameters}, namely $m$, $l$ and $\alpha$ (the first two uniquely determining the constants $r_{+}$ and $r_{-}$ via the relations (\ref{defr+r-})). They can independently be set to \emph{any value}. In particular, it is possible to \emph{set them to zero}, thus immediately obtaining important special subclasses of the spacetime metric (\ref{newmetric}). This is the main advantage of (\ref{newmetric}) if compared to the original form (\ref{metrikaChngMannStelea}) in which, in particular, it is not possible to set ${\alpha=0}$, and also the NUT parameter is not explicitly identified.

Let us now investigate the spacetime, based on the new form of its metric (\ref{newmetric}), (\ref{newfunctions}).

\subsection{The case ${l=0}$: The $C$-metric (accelerating black holes)}

For ${l=0}$ the constants (\ref{defr+r-}) become
\begin{eqnarray}
r_{+} = 2 m\,, \qquad \qquad  r_{-} = 0 \,,  \label{r+r-Cmetrika}
\end{eqnarray}
so that the metric functions (\ref{newfunctions}) reduce considerably to
\begin{eqnarray}
\Omega(r,\theta) \rovno  1-\alpha \, r \cos\theta      \,,\nonumber\\
P(\theta)        \rovno  1-2\alpha m \cos\theta  \,,\nonumber\\
\Q(r)            \rovno  r(r-2m)(1-\alpha r)(1+\alpha r) \,, \label{l=0:newfunctions}\\
\R^2(r,\theta)   \rovno  r^2 \,. \nonumber
\end{eqnarray}
The metric (\ref{newmetric}) thus simplifies to a diagonal line element
\begin{eqnarray}
\dd s^2 \rovno  \frac{1}{(1-\alpha \, r \cos\theta)^2}\,
   \Bigg[ -Q \, \dd t^2 + \frac{\dd r^2}{Q}
   + r^2 \bigg( \frac{\dd\theta ^2}{P} +P \sin^2\!\theta \,\dd\varphi^2 \bigg)\Bigg] , \label{l=0:newmetric}
\end{eqnarray}
where
\begin{eqnarray}
P  \rovno  1-2\alpha m \cos\theta  \,,\nonumber\\
Q \equiv \frac{\Q}{\R^2} \rovno \Big(1-\frac{2m}{r}\Big)(1-\alpha r)(1+\alpha r) \,. \label{l=0:finalfunctions}
\end{eqnarray}
This is exactly the \emph{$C$-metric expressed in spherical-type coordinates}, see Eqs.~(14.6) and~(14.7) in \cite{GriffithsPodolsky:2009}. As has been thoroughly described in Chapter~14 of \cite{GriffithsPodolsky:2009}, this metric represents the spacetime with a pair of Schwarzschild-like black holes of \emph{mass} $m$ which uniformly accelerate due to the tension of cosmic strings (or struts) located along the half-axes of symmetry at ${\theta=0}$ and/or ${\theta=\pi}$. \emph{Their acceleration is determined by the parameter} $\alpha$. This gives the physical interpretation to the two constant parameters of the solution.

\subsection{The case ${\alpha=0}$: The Taub--NUT metric (twisting black holes)}

Complementarily, it is possible to directly set ${\alpha=0}$ in the metric (\ref{newmetric}). In such a case
the functions (\ref{newfunctions}), using the identities (\ref{identity4}), reduce to simple quadratics
\begin{eqnarray}
\Omega(r,\theta) \rovno  1  \,,\nonumber\\
P(\theta)        \rovno  1  \,,\nonumber\\
\Q(r)            \rovno  \big(r-r_{+}\big)\big(r-r_{-}\big) \equiv r^2 -2m r -l^2 \,, \label{alpha=0:newfunctions}\\
\R^2(r,\theta)   \rovno  r^2 + l^2\,. \nonumber
\end{eqnarray}
The metric (\ref{newmetric}) remains non-diagonal, but has a compact explicit form
\begin{eqnarray}
\dd s^2 \rovno - f\, \big(\dd t -2l \cos \theta \,\dd\varphi \big)^2 + \frac{\dd r^2}{f}
  + (r^2+l^2)(\dd\theta^2 + \sin^2 \!\theta \,\dd\varphi^2) \,, \label{alpha=0:newmetric}
\end{eqnarray}
where
\begin{eqnarray}
f \equiv \frac{\Q}{\R^2} \rovno \frac{r^2 - 2mr - l^2}{r^2+l^2} \,. \label{alpha=0:finalfunction}
\end{eqnarray}
It is exactly the \emph{standard Taub-NUT metric}, see Eqs. (12.1) and (12.2) in \cite{GriffithsPodolsky:2009}. As summarized in Chapter~12 of \cite{GriffithsPodolsky:2009}, this metric is interpreted as a spacetime with black hole of \emph{mass} $m$ and \emph{NUT twist parameter} $l$. There are horizons located at ${r=r_{+}}$ and ${r=r_{-}}$, but there is \emph{no curvature singularity} at ${r=0}$. Whenever the NUT parameter $l$ is nonvanishing, there is an internal twist in the geometry, related to spinning cosmic strings located along the axes  ${\theta=0}$ and/or ${\theta=\pi}$. In the vicinity of these ``torsion singularities'' there appear closed timelike curves.

\subsection{The case ${\alpha=0}$ and ${l=0}$: Schwarzschild black hole}

By simultaneously setting both the acceleration $\alpha$ and the NUT parameter $l$ to zero, we immediately obtain the standard spherically symmetric metric
\begin{eqnarray}
\dd s^2 \rovno - \Big(1-\frac{2m}{r}\Big)\, \dd t^2 + \Big(1-\frac{2m}{r}\Big)^{-1}\dd r^2
  + r^2(\dd\theta^2 + \sin^2 \!\theta \,\dd\varphi^2) \,. \label{alpha,l=0:newmetric}
\end{eqnarray}
 As is well known (see, e.g. Chapter~8 of \cite{GriffithsPodolsky:2009}), it represents the \emph{spherically symmetric Schwarzschild black hole of mass} $m$ in asymptotically flat space. There is no acceleration and no twist, the axes are regular (there are no cosmic strings, struts, or torsion singularities).

\subsection{The case ${\alpha=0}$ and ${l=0}$ and ${m=0}$: Minkowski flat space}

By setting ${\alpha=0=m}$ in (\ref{l=0:newmetric}), (\ref{l=0:finalfunctions}) which implies ${P=1=Q}$, or by setting ${l=0=m}$ in (\ref{alpha=0:newmetric}), (\ref{alpha=0:finalfunction}) which implies ${f=1}$, or by setting ${m=0}$ in (\ref{alpha,l=0:newmetric}), we obtain
\begin{eqnarray}
\dd s^2 \rovno - \dd t^2 + \dd r^2 + r^2(\dd\theta^2 + \sin^2 \!\theta \,\dd\varphi^2) \,. \label{Minkowski}
\end{eqnarray}
This is obviously the flat metric in spherical coordinates (Eq. (3.2) in \cite{GriffithsPodolsky:2009}).

Since all such subcases are \emph{directly} obtained as special cases, \emph{it is indeed natural to interpret the general metric} (\ref{newmetric}), (\ref{newfunctions}) \emph{as} a three-parameter family of exact spacetimes with \emph{uniformly accelerating black holes with the twist NUT parameter}.

\newpage

\begin{figure}[h!]
\centerline{\includegraphics[scale=1]{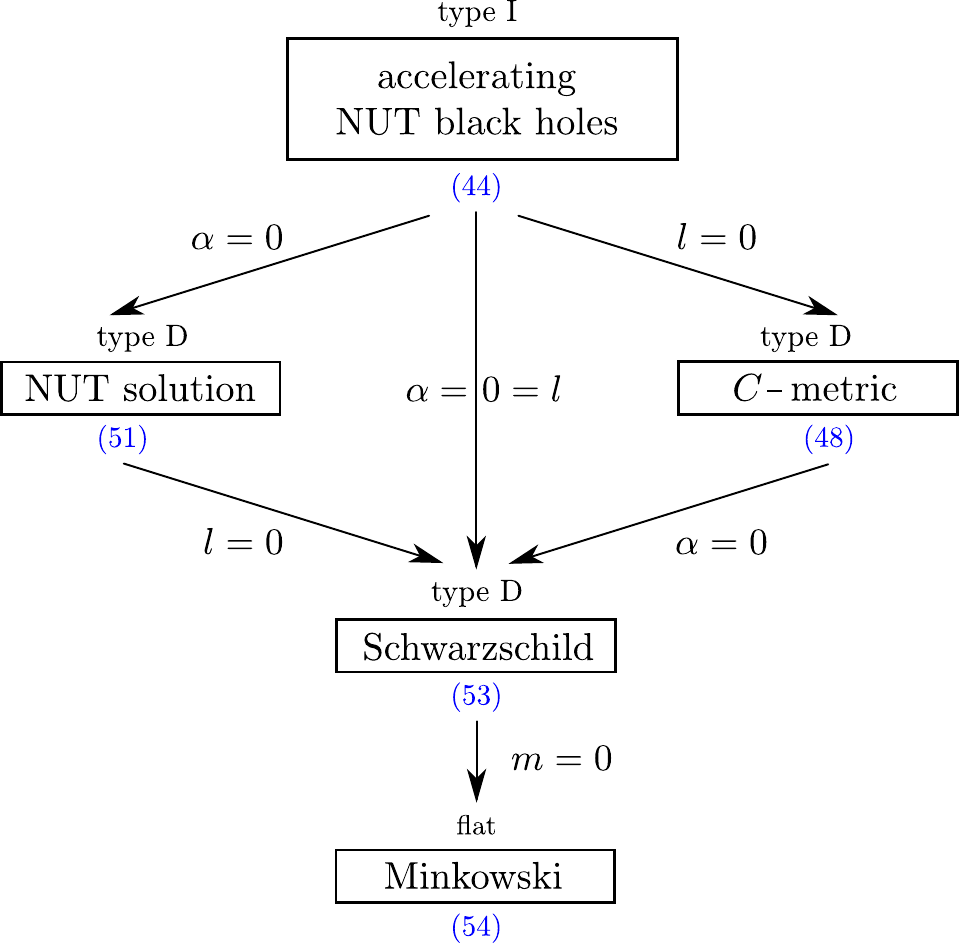}}
\vspace{4mm}
\caption{ \small Schematic structure of the complete family of accelerating black holes with a NUT parameter. This 3-parameter class of vacuum solutions to Einstein's field equations is of general algebraic type~I, reducing to double degenerate type~D whenever the acceleration $\alpha$ or the NUT parameter $l$ (or both) vanish. By setting any of the three independent parameters $\alpha$, $l$, $m$ to zero, the well-known classes (namely the NUT solution, the $C$-metric, Schwarzschild black hole and Minkowski flat space) are obtained directly in their usual forms, whose equation numbers are also indicated in the diagram.}
\label{diagram}
\end{figure}

The structure of the new family of spacetimes which represent accelerating NUT black holes is shown in Fig.~\ref{diagram}. Previously known spacetimes are obtained in their classic form  by simply setting the acceleration $\alpha$, the NUT parameter $l$, or the mass $m$ to zero. With these settings, algebraically general solution of Einstein's vacuum equations reduces to type~D.

\vspace{4mm}
\

\newpage

\section{Physical interpretation of the new metric form}
\label{interpretation}

\subsection{Position of the horizons}
\label{subsec:horizon}

The metric (\ref{newmetric}) is very convenient for investigation of horizons. In these coordinates, $\partial_t$ is one of the Killing vectors (the second  is $\partial_\varphi$). Its norm is ${-\Q /(\Omega\R)^{2}}$, so that $t$ is a temporal coordinate in the regions where ${\Q(r)>0}$, while it is a spatial coordinate in the regions where ${\Q(r)<0}$. These regions are separated by the \emph{Killing horizons} localized at ${\Q(r)=0}$. The form of the metric function~$\Q$ is given by (\ref{newfunctions}), which is clearly a \emph{quartic factorized into four roots}. There are thus \emph{four Killing horizons}, located at
\begin{eqnarray}
\HH_b^+: \qquad r \rovno  r_b^+ \equiv r_{+} >0  \,,\nonumber\\
\HH_b^-: \qquad r \rovno  r_b^- \equiv r_{-} <0 \,,\nonumber\\
\HH_a^+: \qquad r \rovno  r_a^+ \equiv r_{-} + \alpha^{-1}\,, \label{horizons}\\
\HH_a^-: \qquad r \rovno  r_a^- \equiv r_{-} - \alpha^{-1}\,, \nonumber
\end{eqnarray}
(see Fig.~\ref{globalstructure}) where $r_{\pm}$ are defined by (\ref{defr+r-}). Recall also (\ref{identity4}), that is ${\,r_{+} - r_{-} = 2\sqrt{m^2+l^2} >0\,}$ (unless ${m=0=l}$, in which case ${r_{+} =0= r_{-}}$).

The horizons ${\HH_b^+, \HH_b^-}$ at ${r_b^+, r_b^-}$ are \emph{two black-hole horizons}. Interestingly, they are located \emph{at the same values} ${r_{+}, r_{-}}$ of the radial coordinate $r$ as the two horizons in the standard (non-accelerating) \emph{Taub--NUT metric}, see (\ref{alpha=0:newfunctions}).

The horizons ${\HH_a^+, \HH_a^-}$ at ${r_a^+, r_a^-}$ are \emph{two acceleration horizons}. Their presence is the consequence of the fact that the black hole accelerates whenever the parameter ${\alpha}$ is non-zero. They generalize the acceleration horizons ${+ \alpha^{-1}, - \alpha^{-1}}$ present in the \emph{$C$-metric}, see (\ref{l=0:finalfunctions}).

These pairs of roots are clearly ordered as ${r_b^+>r_b^-}$ and ${r_a^+>r_a^-}$ (naturally assuming that the acceleration parameter $\alpha$ is positive). Their mutual relations, however, depend on the specific values of the three physical parameters ${m, l, \alpha}$. Concentrating on the physically most plausible case when the \emph{acceleration is small}, the value of $\alpha^{-1}$ is very large, and $r_a^+$ becomes bigger than~$r_b^+$.
This condition ${r_a^+>r_b^+}$ explicitly reads
\begin{equation}
\alpha < \frac{1}{2\sqrt{m^2+l^2}} \,. \label{orderofhorizonsexpl}
\end{equation}
For such a small acceleration of the black hole, the ordering of its four horizons is
\begin{equation}
r_a^-<r_b^-<0<r_b^+<r_a^+\,. \label{orderofhorizons}
\end{equation}

The first two horizons $\HH_a^-$ and $\HH_b^-$ (acceleration and black-hole, respectively) are in the region ${r<0}$, while the remaining two horizons $\HH_b^+$ and $\HH_a^+$  (black-hole and acceleration, respectively) are in the region ${r>0}$. Such a situation can be naturally understood as the Taub-NUT spacetime with usual two ``inner''  black hole horizons $\HH_b^\pm$, which are here surrounded by two additional ``outer'' acceleration horizons $\HH_a^\pm$ (one in the region ${r>0}$ and the second in the region ${r<0}$).

Evaluating ${\Q(r)}$, generally given by (\ref{newfunctions}), at ${r=0}$ we obtain using (\ref{identity4})
\begin{equation}
\Q(r=0) =  r_{+}r_{-}\big(1-\alpha^2 r_{-}^2\big)
    =  - l^2\big(1-\alpha^2 r_{-}^2\big) \,. \label{Q(0)}
\end{equation}
From the condition (\ref{orderofhorizonsexpl}) and (\ref{defr+r-}) it follows that
\begin{equation}
1-\alpha^2 r_{-}^2 > \frac{2m^2+ 3l^2 +2m \sqrt{m^2+l^2}}{4 (m^2+l^2)}>0 \,, \label{auxil1}
\end{equation}
so that ${\Q(r=0)<0}$. It implies ${\Q<0}$ for any ${r\in (r_b^-, r_b^+)}$. We conclude that the coordinate~$t$ is \emph{temporal} in the regions $(r_b^+,r_a^+)$ and  $(r_a^-,r_b^-)$, that is \emph{between} the black-hole and acceleration horizons, while it is spatial in the complementary three regions of the radial coordinate~$r$.

Moreover, when the condition (\ref{orderofhorizonsexpl}) is satisfied, the metric coefficient $P(\theta)$ in (\ref{newmetric}) is \emph{always positive}. Indeed,
\begin{equation}
P_{\rm min}=P(\theta=0) = 1-\alpha \, (r_{+} - r_{-})
= 1-2\alpha \sqrt{m^2+l^2}>0 \,. \label{Pmin>0}
\end{equation}

Of course, for other choices of the physical parameters, \emph{different number} and \emph{different ordering} of the horizons can be achieved. They also may coincide, thus becoming \emph{degenerate horizons}. In particular, in the limit of \emph{vanishing acceleration} ${\alpha\to0}$, the two outer acceleration horizons disappear (formally via the limits ${r_a^+\to+\infty }$, ${r_a^-\to-\infty }$), and only two Taub-NUT black hole horizons ${\HH_b^+, \HH_b^-}$ remain. On the other hand, for \emph{vanishing NUT parameter} ${l\to0}$, one of the black-hole horizon disappears (formally via the limit ${r_b^-\equiv r_{-}\to0}$), while the second becomes ${r_b^+\equiv r_{+}\to 2m}$. There is just one black-hole horizon at $2m$ surrounded by two acceleration horizons located at $\pm \alpha^{-1}$, which is exactly the case of the $C$-metric with a curvature singularity at ${r=0}$.

\subsection{Curvature of the spacetime, algebraic structure and regularity}
\label{subsec:curvature}

\subsubsection{The Weyl scalars}
We now employ the Weyl scalars $\Psi_A$  given by (\ref{eq:Psi}), (\ref{K}), (\ref{XiEXPLICIT}) to discuss the algebraic properties of the spacetime, including the subcases ${l=0}$ and ${\alpha=0}$, the location of physical curvature singularities and its global structure.

These scalars correspond to the metric (\ref{metricAC}) with coordinates $x,y$, and it is thus natural to denote them as $\Psi_A^{(xy)}$. It will also be convenient to express these curvature scalars as $\Psi_A^{(r\theta)}$ for the metric form (\ref{newmetric}) with coordinates $r,\theta$. Using the transformation (\ref{transformation_to_newform}) and definitions (\ref{F->F}), (\ref{newfunct}) we immediately derive ${\alpha^2 (1-x^2)F(x) (y^2-1)F(y) = P\Q\,(r-r_{-})^{-4}\, \sin^2\theta}$, with ${P=P(\theta)}$ and ${\Q=\Q(r)}$ given by (\ref{newfunctions}), and similarly we express the functions $\Xi$, $\Sigma$ and $\Pi$.
However, it is \emph{also necessary to properly rescale} the scalars $\Psi_A^{(xy)}$ given by (\ref{eq:Psi}) to get $\Psi_A^{(r\theta)}$ because the metrics (\ref{metricAC}) and (\ref{newmetric}) are \emph{not} the same: They are related by a \emph{constant conformal factor},
\begin{equation}
g_{ab}^{(r\theta)} = \omega^2 \,g_{ab}^{(xy)}\,,\qquad \hbox{where}\quad
  \omega^2 = \frac{r_{+}}{r_{+}-r_{-}}\,. \label{scalingomega}
\end{equation}
Indeed, ${g_{ab}^{(xy)} = 2\, \bar{g}_{ab} }$ while ${g_{ab}^{(r\theta)} = 2\,\frac{r_{+}}{r_{+}-r_{-}} \,\bar{g}_{ab} }$, see (\ref{rescaling}).
The corresponding Weyl tensor components are related as ${C_{abcd}^{(r\theta)}=\omega^2 \, C_{abcd}^{(xy)}}$, see \cite{Wald:book1984}. The null tetrad (\ref{tetrad}) also needs to be rescaled in such a way that it remains properly normalized in the coordinates $r,\theta$ as (\ref{nulnorm}). This requires
${\boldk^{(r\theta)}=\omega^{-1}\boldk^{(xy)}}$, ${\boldl^{(r\theta)}=\omega^{-1}\boldl^{(xy)}}$, ${\boldm^{(r\theta)}=\omega^{-1}\boldm^{(xy)}}$. In view of (\ref{defPsi}), we obtain the relation
\begin{equation}
\Psi_A^{(r\theta)}= \omega^{-2} \,\Psi_A^{(xy)} \,. \label{scalingPsi}
\end{equation}
Using (\ref{eq:Psi})--(\ref{XiEXPLICIT}) and (\ref{scalingomega})--(\ref{scalingPsi}), we thus calculate the Weyl curvature scalars for the metric (\ref{newmetric}) with respect to the null tetrad
\begin{eqnarray}
\quad
\boldk^{(r\theta)}  \rovno  \frac{1}{\sqrt{2}}\,\Omega\,\bigg( \frac{\R}{\sqrt{\Q}} \,\partial_t + \frac{\sqrt{\Q}}{\R} \,\partial_r \bigg) , \nonumber \\
\boldl^{(r\theta)}  \rovno  \frac{1}{\sqrt{2}}\,\Omega\,\bigg( \frac{\R}{\sqrt{\Q}} \,\partial_t - \frac{\sqrt{\Q}}{\R} \,\partial_r \bigg) ,  \label{tetradnew}\\
\boldm^{(r\theta)} \rovno \frac{1}{\sqrt{2}}\, \frac{\Omega}{\R\, \sqrt{P} \sin \theta}
 \, \Big( \partial_\varphi  + 2l \big(\cos\theta - \alpha\,\T\sin^2\!\theta \,\big)\,\partial_t
 -{\rm i}\, P \sin \theta\,\partial_\theta  \Big) . \nonumber
\end{eqnarray}
It turns out that
\begin{eqnarray}
\Psi_0^{(r\theta)} = \Psi_4^{(r\theta)} \rovno -3\,{\rm i}\, \alpha^2\, l \, P \Q \,  (r-r_{-})\sin ^2 \theta \, X  \,,
  \nonumber\\[0.25cm]
\Psi_1^{(r\theta)} = \Psi_3^{(r\theta)} \rovno 3\, \alpha\, l\,  \sqrt{P\Q} \,\sin \theta  \,S \, X  \,,
  \label{eq:Psi-spher} \\[0.25cm]
\Psi_2^{(r\theta)} \rovno \Big[-r_{+}\sqrt{m^2 + l^2}\, \Omega^5 + {\rm i}\,l \, W/(r-r_{-}) \Big]  \, X \,,
  \nonumber
\end{eqnarray}
where
\begin{eqnarray}
X (r,\theta) \rovno \frac{(r_{+}^2 + l^2)(r-r_{-})^3\, \Omega^4}{\big[\,  r_{+}(r-r_{-})^2 \, \Omega^2 -{\rm i}\, l\, \Q \big]^3} \,, \nonumber \\
S(r,\theta) \rovno \big(1-\alpha ^2(r-r_{-})^2\big)(r-r_{+}) - \Big[(r-r_{+}) - \sqrt{m^2 + l^2}\, \big(1-\alpha ^2(r-r_{-})^2\big) \Big] \Omega \,, \label{xiws-spher}\\
W(r,\theta) \rovno
2\, S^2 + \big(1-\alpha ^2(r-r_{-})^2\big)(r-r_{+})
  \Big[ \sqrt{m^2 + l^2}\, \Omega^3 - \alpha ^2(r-r_{-})^3 P \sin ^2 \theta \Big]
   \,. \nonumber
\end{eqnarray}
These functions are related to (\ref{K}) via
\begin{equation}
X   \equiv  \frac{-{\rm i}\,(r_{+}-r_{-})}{\alpha^2\,r_{+}^2(r-r_{-})^5}\,\Xi \,, \qquad
S     \equiv  \alpha^2\,(r-r_{-})^3\,\Sigma \,, \qquad
W     \equiv  \alpha^4\,(r-r_{-})^6\,\Pi \,,
 \label{defNoveFce}
\end{equation}
and ${\Omega=\Omega(r,\theta)}$, ${P=P(\theta)}$ and ${\Q=\Q(r)}$ are given by (\ref{newfunctions}).





As we have already argued in Sec.~\ref{sec_algebrtype}, this class of spacetimes with accelerating Taub--NUT black hole is generically of \emph{type~I}, i.e., it is \emph{algebraically general}. However, it may degenerate. When \emph{either} ${\alpha=0}$ \emph{or} ${l=0}$, the only nontrivial curvature component is given by
\begin{equation}
\Psi_2^{(r\theta)} = \Big[-r_{+}\sqrt{m^2 + l^2}\, \Omega^5 + {\rm i}\,l \, W/(r-r_{-}) \Big]  \, X  \,.
 \label{Psi2TypeD}
\end{equation}
Such spacetimes are clearly of \emph{algebraic type~D}, with two double-degenerate principal null directions $\boldk^{(r\theta)}$ and $\boldl^{(r\theta)} $ of the Weyl/Riemann tensor.

This is fully consistent with the fact that the case ${l=0}$ (implying ${r_{+} = 2 m}$, ${r_{-} = 0}$, see (\ref{r+r-Cmetrika}), and ${X=(r_{+}\, r^3 \, \Omega^2)^{-1}}$) corresponds to the type~D \emph{accelerating $C$-metric}, for which
\begin{equation}
\Psi_2^{(r\theta)} = -\frac{m}{r^3}\,(1-\alpha \, r \cos\theta)^3 \,, \label{Psi2Cmetric}
\end{equation}
see Chapter~14 in \cite{GriffithsPodolsky:2009}.

The complementary case ${\alpha=0}$, which cannot be directly obtained from $\Psi_A^{(xy)}$ given by (\ref{eq:Psi}), corresponds to the type~D \emph{twisting Taub--NUT metric}. It follows form (\ref{alpha=0:newfunctions}) that in such a case ${\Omega=1}$ and ${\Q(r)=(r-r_{+})(r-r_{-})}$. With the help of relation (\ref{identity4}) we thus get
\begin{eqnarray}
X  \rovno \frac{r_{+}^2 + l^2}{\big[\,  r_{+}(r-r_{-})  -{\rm i}\, l\, (r-r_{+}) \big]^3}
 = \frac{(r_{+} -{\rm i}\, l)(r_{+} +{\rm i}\, l)}{(r_{+}-{\rm i}\, l)^3(r +{\rm i}\, l)^3} \,, \nonumber \\[2mm]
S \rovno \sqrt{m^2 + l^2}   \,, \qquad
W = \sqrt{m^2 + l^2}\,(r-r_{-})  \,,\label{xiws-spherNUT}
\end{eqnarray}
so that
\begin{eqnarray}
\Psi_2^{(r\theta)} \rovno -\sqrt{m^2 + l^2}\, (r_{+} - {\rm i}\,l )  \, X
= -\frac{\sqrt{m^2 + l^2}}{r_{+}-{\rm i}\, l} \,\frac{r_{+} +{\rm i}\, l}{(r +{\rm i}\, l)^3}
= -\frac{\sqrt{m^2 + l^2} }{r_{+}^2+ l^2} \,\frac{(r_{+} +{\rm i}\, l)^2}{(r +{\rm i}\, l)^3}  \,.
 \label{Psi2TypeDNUT}
\end{eqnarray}
Applying the identities
\begin{equation}
r_{+}^2+ l^2 = r_{+}(r_{+} - r_{-}) = 2r_{+}\sqrt{m^2+l^2}\,,\quad\hbox{and}\quad
(r_{+} +{\rm i}\, l)^2 = 2r_{+}(m +{\rm i}\, l) \,,
 \label{Psi2TypeDNUTpomoc}
\end{equation}
we finally obtain
\begin{equation}
\Psi_2^{(r\theta)} =
    -    \frac{m + {\rm i}\,l}{\,(r+{\rm i}\,l)^{3}} \,, \label{Psi2NUTb}
\end{equation}
which is the standard form of the scalar $\Psi_2$ for the Taub--NUT spacetime, see Chapter~12 in \cite{GriffithsPodolsky:2009}.

\subsubsection{Algebraic type and regularity of the horizons}
It can be immediately observed from (\ref{eq:Psi-spher}) that on the horizons (\ref{horizons}), defined by ${\Q=0}$, all the Weyl scalars vanish except
\begin{equation}
\Psi_2^{(r\theta)}(\hbox{at any horizon } r_h) =
-\frac{2\sqrt{m^2+l^2} }{ r_{+}^2(r_h-r_{-})^3}\,
\Big[r_{+}\sqrt{m^2 + l^2}\, \Omega^3 - {\rm i}\,l \, \frac{W}{(r_h-r_{-})\,\Omega^2} \Big]   \,.
 \label{Psi2TypeDhorizon}
\end{equation}
Therefore,  \emph{all horizons are of algebraic type~D}. This is true in \emph{a generic case with any acceleration~$\alpha$ and any NUT parameter~$l$}. Moreover, at these horizons the spacetime is \emph{regular}, that is \emph{free of curvature singularities}. This can be proved as follows:

\vspace{4mm}
\noindent
$\bullet$ At the \emph{acceleration horizons} ${r_a^+,r_a^-}$, the values are  ${r_h-r_{-}=  \pm \alpha^{-1}}$, so that ${\Omega(r_h) = 1 \mp \cos\theta }$ and
${W(r_h)=2\alpha^{-2}(1 \mp2\alpha\sqrt{m^2+l^2})^2\,  \Omega^2}$, implying
\begin{equation}
\Psi_2^{(r\theta)}(\HH_a^\pm) = 2\alpha^2\,\frac{\sqrt{m^2+l^2}}{ r_{+}^2 } \,
\Big[\mp \alpha\, r_{+}\sqrt{m^2 + l^2}\, (1 \mp \cos\theta)^3
+ 2\,{\rm i}\, l \,(1 \mp 2\alpha\sqrt{m^2+l^2})^2\,  \Big] \,.
 \label{Psi2TypeDhorizonAccel}
\end{equation}
$\bullet$ At the \emph{positive black hole horizon} ${r_b^+\equiv r_{+}>0}$, the value of the factor is  ${r_h-r_{-}=2\sqrt{m^2+l^2}}$, so that ${\Omega(r_h) = 1-2\alpha\sqrt{m^2+l^2} \cos\theta = P}$, ${W(r_h) =  2\,(m^2 + l^2) \big(1-4\alpha ^2(m^2+l^2)\big)^2 \, \Omega^2}$. Thus,
\begin{equation}
\Psi_2^{(r\theta)}(\HH_b^+) =
-\frac{1}{ 4\,r_{+}^2\sqrt{m^2+l^2}}\,
\Big[\,r_{+}\big(1-2\alpha\sqrt{m^2+l^2} \cos\theta\big)^3
- {\rm i}\,l \, \big(1-4\alpha ^2(m^2+l^2)\big)^2 \Big]   \,.
 \label{Psi2TypeDhorizonBH1}
\end{equation}
$\bullet$ At the \emph{negative black hole horizon} ${r_b^-\equiv r_{-} <0}$, the expression (\ref{Psi2TypeDhorizon}) seems to diverge. However, a careful analysis of the limit ${r \to r_{-}}$ of (\ref{Psi2TypeD}) shows, using ${X \to {\rm i}\, r_{+}\big(4l^3 (m^2 + l^2)\big)^{-1}}$, ${\Omega \to 1}$ and
${W/ (r_h-r_{-}) \to \sqrt{m^2+l^2}\,\big(1-6\alpha\sqrt{m^2+l^2} \cos\theta\big)}$ that
\begin{equation}
\Psi_2^{(r\theta)}(\HH_b^-) =
-\frac{r_{+}}{4\,l^3\, \sqrt{m^2 + l^2}}\Big[\,l \, \big(1-6\alpha\sqrt{m^2+l^2} \cos\theta\big) + {\rm i}\,r_{+}  \Big]   \,.
 \label{Psi2TypeDhorizonBH2}
\end{equation}
The expressions (\ref{Psi2TypeDhorizonAccel})--(\ref{Psi2TypeDhorizonBH2}) explicitly demonstrate that \emph{at any horizon the gravitational field is finite}, without the curvature singularities.

\subsubsection{Algebraic type of the axes and principal null directions}
Similarly, \emph{along both the axes} ${\theta=0}$ and ${\theta=\pi}$ the function $\sin\theta$ vanishes, which implies that ${\Psi_0^{(r\theta)}=\Psi_1^{(r\theta)}=0=\Psi_3^{(r\theta)}=\Psi_4^{(r\theta)}}$. This proves that the \emph{algebraic structure of the spacetime on theses axes is also of type~D}, with the only curvature component (\ref{Psi2TypeD}).

Finally, let us comment on the \emph{principal null directions} (PNDs) of the curvature tensor introduced in Sec.~\ref{PND}. Using the Weyl scalars (\ref{eq:Psi-spher}) we can express the key discriminant of the equation (\ref{kvadrsol}) as
\begin{equation}
\disc  \equiv 4\Psi_1 ^2 -2\Psi_0(3\Psi_2-\Psi_0) = -18\, \alpha^2\, l\, \sqrt{m^2+l^2}\, P\Q\,\sin^2\theta\, \Omega^3 X^2 \,Y  \,,
\label{discrim}
\end{equation}
where ${Y(r,\theta)=l\,\big(1-\alpha ^2(r-r_{-})^2\big)(r-r_{+}) + {\rm i}\,r_{+}(r-r_{-}) \, \Omega^2}$. Therefore, there are \emph{in general two distinct roots} ${\kappa_1,\kappa_2}$ of (\ref{kvadrsol}),  and subsequently there are \emph{four distinct roots $K_i$} of (\ref{kvartrsol}). They correspond to \emph{four distinct PNDs} of the Weyl tensor, confirming that the metric (\ref{newmetric}) is of algebraically general type~I.

However, if (and only if) ${\alpha=0}$ or ${l=0}$, the discriminant (\ref{discrim}) \emph{everywhere vanishes} and there is \emph{only one double root} $\kappa$ of (\ref{kvadrsol}). In such cases, there are just two roots
\begin{eqnarray}
K_{1, 2} = \frac{\kappa \pm \sqrt{\kappa ^2 -4}}{2} \,,
\label{kvartrsol2}
\end{eqnarray}
corresponding to two \emph{doubly degenerate PNDs} ${\boldk'_{1,2}}$ of type~D spacetimes (the Taub--NUT metric and the $C$-metric, respectively). In particular, in this limit ${K_{1} \to 0}$ and ${K_{2} \to \infty}$ which effectively corresponds to PND $\boldk^{(r\theta)}$ and PND $\boldl^{(r\theta)}$ given by (\ref{tetradnew}).

\subsection{Curvature singularities and invariants}
\label{subsec:singularity}

\subsubsection{Investigation of possible singularities}
The Weyl scalars $\Psi_A^{(xy)}$ given by (\ref{eq:Psi})--(\ref{XiEXPLICIT}), or their equivalent forms  $\Psi_A^{(r\theta)}$ given by (\ref{eq:Psi-spher})--(\ref{xiws-spher}), can be used to study \emph{curvature singularities} in the family of accelerating NUT black holes.

By inspection we observe that all functions entering these scalars are bounded\footnote{As will be demonstrated in Sec.~\ref{subsection:confinity}, a possible divergence for ${r\to\infty}$ corresponds to asymptotically flat regions.} \emph{except} the function $X(r,\theta)$, or equivalently $\Xi(x,y)$, whose denominator can be zero. This key function appears as a \emph{joint factor in all the Weyl scalars} (\ref{eq:Psi-spher}). Regions of spacetime where ${X(r,\theta)\to\infty}$ thus clearly indicate the possible presence of a physical singularity. In view of (\ref{xiws-spher}), such a \emph{curvature singularity} corresponds to the vanishing denominator of $X$ (provided its numerator remains nonzero), that is
\begin{equation}
r_{+}(r-r_{-})^2 \, \Omega^2 -{\rm i}\, l\, \Q =0 \,.  \label{X-singular}
\end{equation}
\emph{Both} the real and imaginary parts must vanish. Since ${r_{+} = m + \sqrt{m^2+l^2}>0}$, $\Omega$ is everywhere a positive conformal factor, and ${\Q=0}$ identifies regular horizons (as shown in previous section), the only possibility is when
\begin{equation}
l=0 \qquad\hbox{and at the same time}\qquad r=r_{-}=0 \,,  \label{singularity}
\end{equation}
where in the last equality we applied the relation ${r_{-} \equiv  m - \sqrt{m^2+l^2}}$ for ${l=0}$. The curvature singularity thus appears \emph{only} in the $C$-metric spacetime at the origin ${r=0}$. \emph{All other spacetimes in the large class of accelerating NUT black holes are nonsingular}. The presence of the NUT parameter $l$ (even a very small one) thus makes the spacetime regular. This property is well known for classic Taub--NUT spacetime (see Chapter~12 in \cite{GriffithsPodolsky:2009}), and the same property holds also in this new class of accelerating NUT black holes. Consequently, to describe the complete spacetime manifold, it is necessary to consider the \emph{full range of the radial coordinate} ${r\in(-\infty,+\infty)}$.

To confirm these observations, we employ the \emph{scalar curvature invariant} $I$  defined in (\ref{IJ}). Introducing a convenient new function $\Delta$, defined as
\begin{equation}
\Delta \equiv  \Psi_2 - \Psi_0   \,, \label{Delta}
\end{equation}
and using the special geometrical property of the spacetime ${\Psi_0 = \Psi_4}$ and ${\Psi_1 = \Psi_3 }$, this invariant is simplified to
\begin{eqnarray}
I \rovno \Psi_0^2-4 \Psi_1^2 + 3\Psi_2 ^2 = 3 \Delta^2 - \disc  \,,  \label{Iagain}
\end{eqnarray}
where the discriminant $\disc$ is given by (\ref{discrim}). Explicit evaluation now leads to
\begin{eqnarray}
I \rovno 3 \,\Big(\,r_{+}^2(m^2+l^2) \, \Omega^{10}  - 12 \, \alpha^2 \, l^2 P\Q\,\sin ^2\!\theta\, S^2
- 3 \, \alpha^4 l^2 (r-r_{-})^2 P^2\Q^2\sin ^4 \theta     \nonumber \\
&& \hspace{15mm} - l^2 \, W^2 /(r-r_{-})^2 -2  \,  {\rm i} \,  l \,r_{+}\sqrt{m^2+l^2}\, \Omega^5 \, W /(r-r_{-})  \Big) \, X^2 \,.
\end{eqnarray}
Since (as already argued)  even the function ${W/(r-r_{-})}$ is finite at the black hole horizon ${r_b^-\equiv r_{-}}$, the scalar curvature invariant $I$ becomes unbounded only if the function $X$ diverges. This happens if, and only if, both the conditions (\ref{singularity}) hold.

Recall also that the real part of the invariant $I$ is proportional to the \emph{Kretschmann scalar},
\begin{equation}
{\cal K}\equiv R_{abcd}\,R^{abcd} = 16\,{\cal R}e\,(I)\,,  \label{Kretschmann}
\end{equation}
which can thus be evaluated as
\begin{eqnarray}
{\cal K} = 48\,\Big\{{\cal R}e\,(\Psi_2^2)-3 \, \alpha^2  l^2 P\Q  \sin^2 \theta \, \big[4S^2+\alpha^2 (r-r_{-})^2 P\Q \sin^2 \theta \big]\, {\cal R}e\,(X^2)\Big\} \,.   \label{KretschmannEXPLIC}
\end{eqnarray}
In this form it is explicitly seen that the Kretschmann scalar for the $C$-metric or the Taub--NUT black hole is simply obtained by  setting  ${l=0}$ or ${\alpha=0}$, respectively. In both cases, it leads  to
\begin{equation}
{\cal K}_{l\,{\rm or}\,\alpha \rightarrow 0} = 48 \, {\cal R}e\,(\Psi^2_2) \,,
\end{equation}
where $\Psi_2$ is given by (\ref{Psi2Cmetric}) or (\ref{Psi2NUTb}), in full agreement with \cite{GuhaChakraborty:2019, ClementGaltsovGuenouche:2016}. Interestingly, ${{\cal K}=48\,{\cal R}e\,(\Psi^2_2)}$ also on the horizons (\ref{horizons}) where ${\Q=0}$, and on the axes ${\theta=0,\pi}$ where $\sin\theta=0$.

\newpage
In the general case of accelerating NUT back holes, the Kretschmann curvature scalar ${\cal K}$ is given by expression~(\ref{KretschmannEXPLIC}). This explicit but somewhat complicated function of the coordinates $r$ and $\theta$ is visualized in the following two figures.

\begin{figure}[ht!]
\centerline{\includegraphics[scale=0.30]{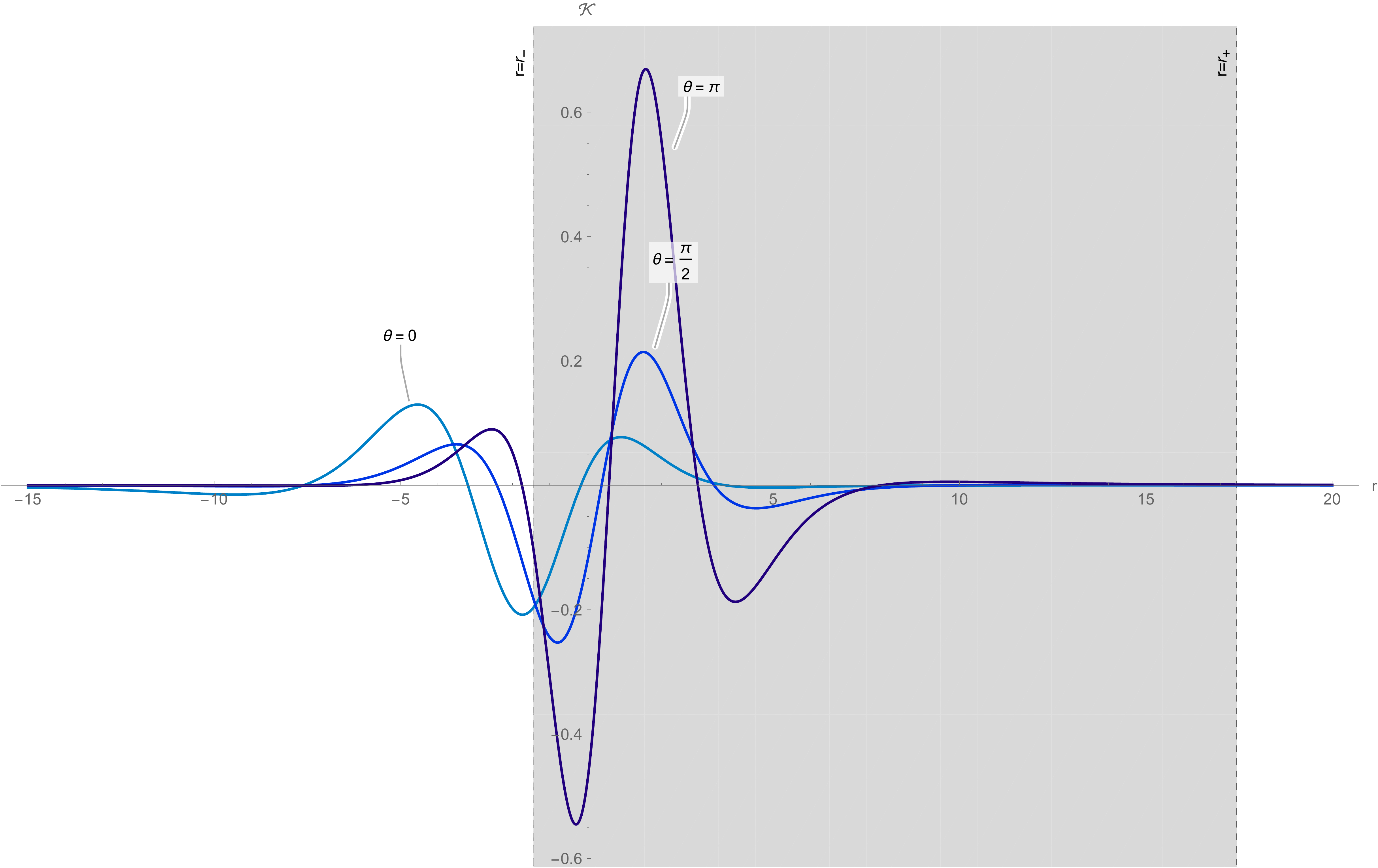}}
\vspace{2mm}
\caption{\small
The value of the Kretschmann curvature scalar (\ref{KretschmannEXPLIC}) plotted as the function ${\cal K}(r)$, where $r$~is the radial coordinate, for ${\theta = 0}$, ${\frac{\pi}{2}}$, and $\pi$. The black-hole parameters are ${m=8}$, ${l=5}$, and ${\alpha=0.025}$.}
\label{figureKretschmann2D}
\end{figure}

In~Fig.~\ref{figureKretschmann2D} we plot the Kretschmann scalar ${\cal K}(r)$ as a function of the radial coordinate~$r$ for \emph{three fixed privileged values of} $\theta$, namely ${\theta = 0}$, ${\theta =\frac{\pi}{2}}$ and ${\theta =\pi}$. In fact, we will argue later that the two poles/axes at ${\theta = 0}$ and~$\pi$ correspond to the position of (rotating) cosmic strings, while ${\theta = \frac{\pi}{2}}$ is the equatorial section ``perpendicular'' to them. It can be seen that for each $\theta$ there are several local maxima and local minima. Half of these extremes are in the region ${r>0}$, the remaining are located in the region ${r<0}$. The \emph{curvature is everywhere finite}, and its \emph{maximal values are localized close to the origin ${r=0}$ inside the black hole}, that is within the shaded region ${r\in(r_{-}, r_{+}) \equiv (r_b^-, r_b^+)}$.

\vspace{0mm}
\vspace{4mm}

\begin{figure}[ht!]
\centerline{\includegraphics[scale=0.7]{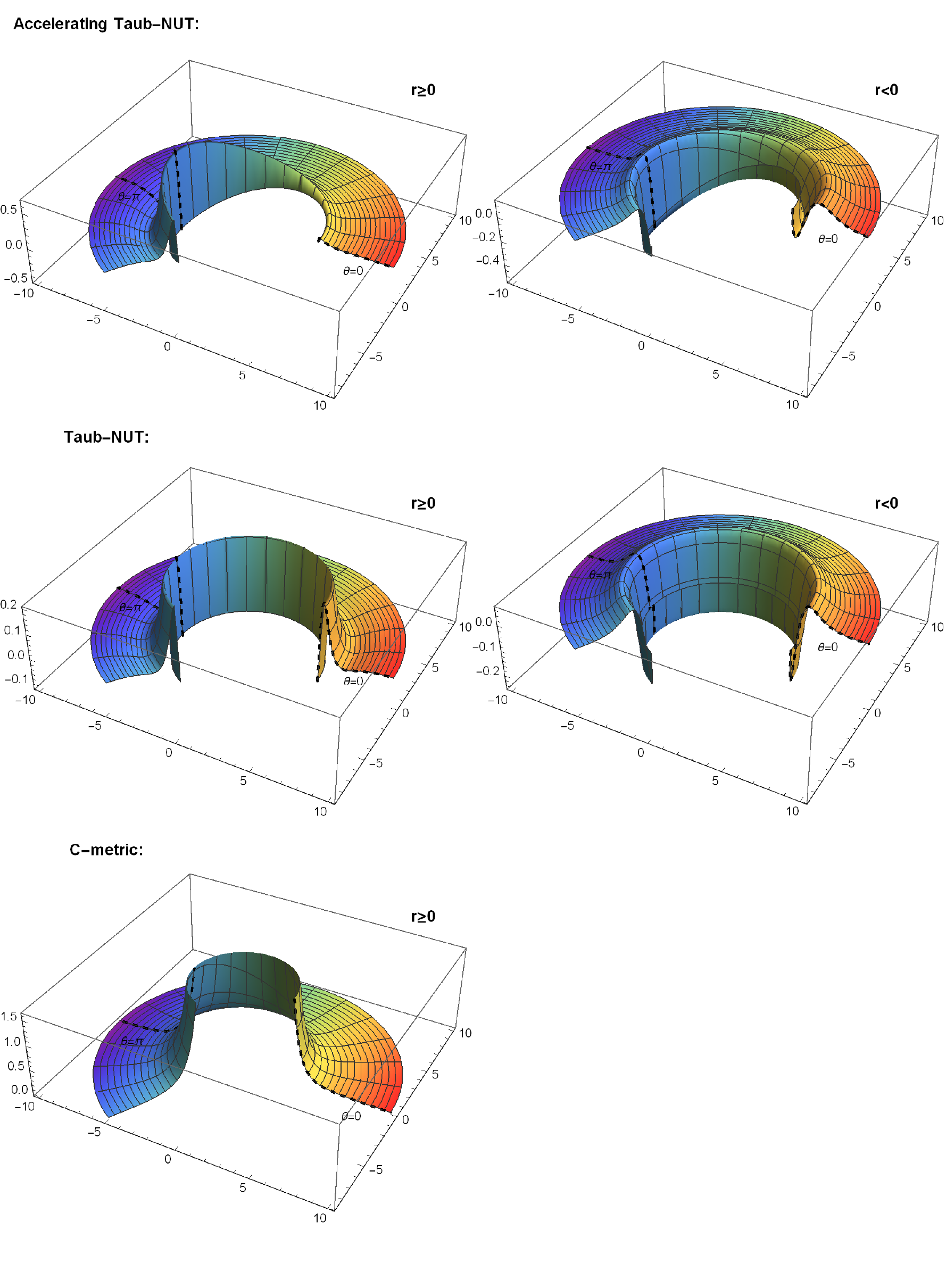}}
\vspace{-2mm}
\caption{\small
The Kretschmann curvature scalar (\ref{KretschmannEXPLIC}) visualized in quasi-polar coordinates as ${{\cal K}({\rm x},{\rm y})}$, where ${{\rm x} \equiv \sqrt{r^2 + l^2}\,\sin \theta}$, ${\mathrm{y} \equiv \sqrt{r^2 + l^2}\,\cos \theta\,}$, so that ${r=0}$ is a circle of radius $l$. The left column  corresponds to ${r\ge0}$, while the right column represents ${r<0}$. The first row plots the Kretchmann scalar for the accelerating NUT black hole with ${m=8}$, ${l=5}$ and ${\alpha=0.025}$. It can be seen that the curvature is everywhere finite, even in the vicinity of ${r=0}$, and it smoothly continues across ${r=0}$ from ${r>0}$ to ${r<0}$. The second and third rows correspond to special cases of this metric, namely the Taub--NUT metric (with ${m=8}$, ${l=5}$, ${\alpha=0}$) and the $C$-metric (with ${m=8}$, ${l=0}$, ${\alpha=0.025}$). The Taub--NUT metric has no divergence of ${\cal K}$, which is independent of $\theta$. On the other hand, the $C$-metric becomes singular as ${r\to0}$, that is at ${{\rm x}=0={\rm y}}$ (therefore we plot only the region ${r\ge0}$). The two separate cosmic strings along the axes  ${\theta = 0}$ and ${\theta =\pi}$ are indicated as dashed curves.}
\label{figureKretschmann}
\end{figure}

In~Fig.~\ref{figureKretschmann} we include the angular dependence on $\theta$. The left column  corresponds to the region ${r\ge0}$, while the right column represents the region ${r<0}$. The first row plots the Kretchmann scalar ${\cal K}(r,\theta)$ for the accelerating NUT black hole (with ${m=8}$, ${l=5}$, ${\alpha=0.025}$), the second and third rows correspond to special cases of this metric, namely the Taub--NUT metric (${m=8}$, ${l=5}$, ${\alpha=0}$) and the $C$-metric (${m=8}$, ${l=0}$, ${\alpha=0.025}$). From these visualizations of the Kretschmann curvature scalar it is seen that the dependence on both $r$ and $\theta$ is smooth, and the curvature is everywhere finite, except for the $C$-metric at ${r=0}$, in full agreement with the condition (\ref{singularity}). The two distinct cosmic strings located on the axes  ${\theta = 0}$ and ${\theta =\pi}$, respectively, are indicated as dashed curves.

\newpage

\subsubsection{Scalar invariants and algebraic types}
Let us conclude this part by returning to the \emph{scalar curvature invariants~$I$ and~$J$}. We can express~$J$, defined in (\ref{IJ}), in terms of the discriminant~$\disc$ and the function $\Delta$ as
\begin{equation}
J = \frac{1}{2} \Delta \big( \disc  - 2 \Delta^2 \big) \,.  \label{Jagain}
\end{equation}
Using (\ref{Iagain}), the key expression ${I^3-27J^2}$ thus takes the compact form
\begin{equation}
I^3-27J^2 = \frac{1}{4} \big( 9 \Delta^2 -4 \disc \big) \disc  ^2 \,,
\end{equation}
which is explicitly
\begin{eqnarray}
I^3-27J^2 \rovno \frac{9}{4} \bigg[ \Big(r_{+} \sqrt{m^2+l^2} \, \Omega^5
- {\rm i} \,  l \, \big[W/(r-r_{-}) - \alpha^2 P\Q \,(r-r_{-})\sin^2\theta	\big] \Big)^2   \nonumber\\
&& \hspace{40mm} -16 \, \alpha^2 l^2  P\Q \sin ^2 \theta \, S^2  \bigg] \disc ^2 X^2 \,.  \label{I2-27J2}
\end{eqnarray}
According to standard classification scheme for determining the algebraic type (see, e.g., page 122 of \cite{Stephanietal:2003}), the spacetime is of a \emph{general algebraic type~I} if (and only if) ${I^3 \ne 27J^2 }$. This is clearly the generic case of (\ref{I2-27J2}), confirming the results of Section~\ref{sec_algebrtype}. Only for ${\disc = 0}$ (or ${X = 0}$ which is, however, a subcase of ${\disc = 0}$), the spacetime degenerates and \emph{becomes algebraically special}. In particular, it follows from (\ref{discrim}) that ${\disc = 0}$ whenever ${\alpha=0}$ or ${l=0}$, and such spacetimes are \emph{actually of type~D everywhere}, as we have already demonstrated in previous sections.

Zeros of the big square bracket in (\ref{I2-27J2}) identify \emph{algebraically more special regions in a given spacetime}. It requires \begin{equation}
W =\alpha^2 P\Q \,(r-r_{-})^2\sin^2\!\theta \qquad\hbox{and}\qquad
r_{+} \sqrt{m^2+l^2} \, \Omega^5 = \pm 4 \, \alpha l  \sqrt{P\Q} \, \sin \theta \, S\,.  \label{specialdegeneracy}
\end{equation}
Clearly, this can happen only for the generic case of accelerating NUT black holes with ${\alpha\ne0\ne l}$. It is interesting to observe from (\ref{eq:Psi-spher}) that these two conditions imply
\begin{equation}
\Psi_2=-\frac{1}{3}(\Psi_0\pm 4\Psi_1) \,,  \label{Psi_2degenregions}
\end{equation}
and thus ${\disc = 4(\Psi_0\pm \Psi_1)^2}$ and ${\Delta = -\frac{4}{3}(\Psi_0\pm \Psi_1)}$, which now implies a specific relation ${\disc=\frac{9}{4}\Delta^2}$. In such degenerate regions, the scalar curvature invariants take the form
\begin{eqnarray}
I \rovno \frac{3}{4}\Delta^2\,, \qquad J = \frac{1}{8}\Delta^3 \,, \qquad\hbox{and further} \nonumber \\[2mm]
K \rovno \frac{9}{8}\Psi_1\,\Delta^2\,,\qquad
L = \frac{1}{4}(\Psi_0\pm3\Psi_1)\,\Delta \quad \Rightarrow \quad   N = \frac{9}{4}\Psi_1(3\Psi_1\pm 2\Psi_0)\,\Delta^2\,, \label{IJdegenregions}
\end{eqnarray}
confirming ${I^3=27J^2}$. Therefore, using the classification scheme, as summarized in~\cite{Stephanietal:2003}, for ${\Delta=0 \Leftrightarrow \Psi_0=\mp \Psi_1}$ the region is of algebraic type~N (because ${I=J=0=K=L}$), while for ${\Delta\ne0}$ it is of type~II. It degenerates to algebraic type~D if, and only if, ${\Psi_1=0\ne\Psi_0}$ (because ${I \ne0\ne J}$ but ${K=0=N}$).

\newpage

\subsection{Description of the conformal infinity~$\scri^\pm$ and global structure}
\label{subsection:confinity}

The coordinates employed in (\ref{newmetric}) are \emph{comoving} in the sense that they are \emph{adapted to the accelerating black holes}. This is clearly seen from the fixed position of the geometrically unique horizons which are \emph{still at the same values} (\ref{horizons}) of the radial coordinate $r$, despite the fact that the black hole moves. This has many advantages, and greatly simplifies physical and geometrical analysis of the spacetime. However, as thoroughly discussed in the simpler case of the $C$-metric (when ${l=0}$) in \cite{GriffithsPodolsky:2009}, such accelerating comoving coordinates can not \emph{naturally cover the whole} conformal infinity~$\scri$~(scri).

\subsubsection{Asymptotically flat regions}
From the Weyl scalars (\ref{eq:Psi-spher}), (\ref{xiws-spher}) it follows that asymptotically flat regions without any curvature, locally resembling the null infinity $\scri$ of Minkowski space, are reached for ${X (r,\theta)\to0}$. It occurs in the vicinity of ${\Omega \equiv  1-\alpha \, (r - r_{-}) \cos\theta   = 0}$, that is for ${r \to r_{-}+1/(\alpha \cos\theta) }$. This corresponds to \emph{the largest possible finite positive} values of~$r$ in the angular half-range ${\theta\in(0,\frac{\pi}{2})}$, but to \emph{the lowest possible finite negative} values of~$r$ for the second half-range ${\theta\in(\frac{\pi}{2},\pi)}$. In the equatorial section ${\theta=\frac{\pi}{2}}$, such asymptotically flat region is reached \emph{both} at ${r=+\infty}$ and ${r=-\infty}$.

\begin{figure}[h!]
\centerline{\includegraphics[scale=1.0]{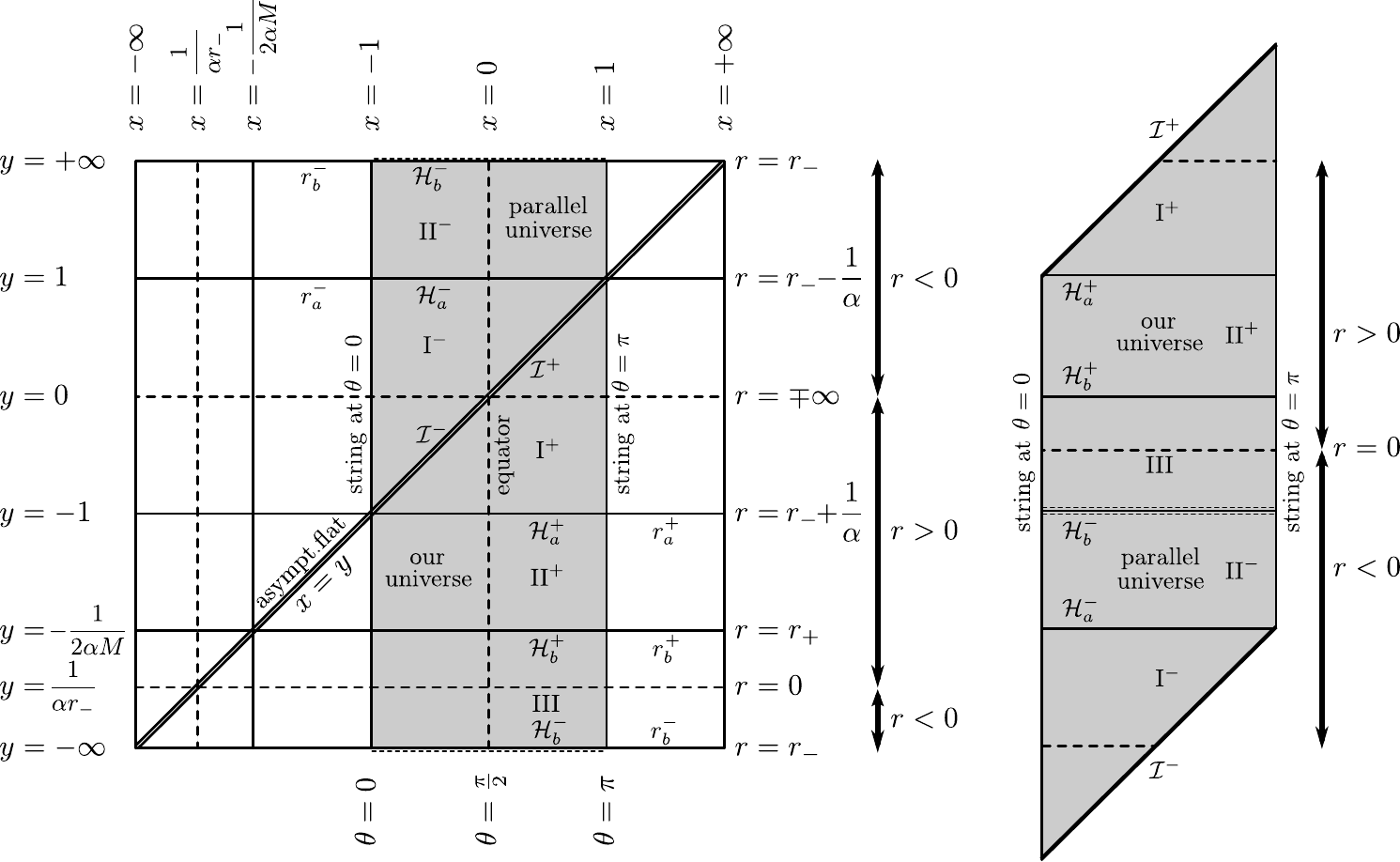}}
\vspace{4mm}
\caption{\small
The complete spacetime structure of the class of accelerating NUT black holes, suppressing the coordinates $t$ and $\varphi$ (corresponding to stationary and axial symmetry). These fundamental sections are represented by (mutually equivalent) coordinates ${x,y}$ and ${\theta,r}$. The black hole spacetime is localized in the shaded region ${x\in[-1,1]}$ between two \emph{rotating cosmic strings} at the two opposite poles ${\theta=0}$ and ${\theta=\pi}$. In the complementary (vertical) direction, the spacetime is separated by four Killing horizons at special values of $y$ and equivalently $r$, namely the two \emph{black-hole horizons} $\HH_b^\pm$ are located at ${r_b^-=r_{-}}$, ${r_b^+=r_{+}}$ and two \emph{acceleration horizons} $\HH_a^\pm$ are at ${r_a^+=r_{-}+\frac{1}{\alpha}}$, ${r_a^-=r_{-}-\frac{1}{\alpha}}$. They separate different regions of the spacetimes in which the coordinate $r$ is spatial (regions~II$^\pm$) or temporal (regions~I$^\pm$ and~III). The values ${r=0}$ and ${r=\mp\infty}$, indicated by horizontal dashed lines, are only coordinate singularities. \emph{Conformal infinity}~$\scri$, where the spacetime is asymptotically flat, is located along the diagonal line ${x=y}$. There are thus \emph{two asymptotically flat regions} corresponding to \emph{``our universe''} where ${r>0}$ and the \emph{``parallel universe''} where ${r<0}$, which are connected through the region III with the highest (but finite) curvature in the black hole interior ${r\in(r_-,r_+)}$. Notice, however, that only along the equatorial section ${\theta=\frac{\pi}{2}}$ the corresponding two conformal infinities $\scri^\pm$ are represented by ${r=\pm\infty}$. Unlike in the  $C$-metric or Schwarzschild black hole, with the NUT parameter $l$ there is \emph{no curvature singularity at} ${r=0}$. It is thus obvious that there are \emph{two complete strings} (not just semi-infinite strings) at ${\theta=0}$ and ${\theta=\pi}$, both connecting the two distinct universes as ${r\in(-\infty,+\infty)}$. In fact, to obtain a geodesically complete spacetime, it is necessary to \emph{``glue the two universes'' along the regular horizon} $\HH_b^-$  at ${r=r_b^-\equiv r_{-}}$, \emph{both at} ${y=-\infty}$ and ${y=+\infty}$, by identifying the corresponding parts of these lower and upper boundaries of the diagram indicated by two finely dashed line segments between ${x\in[-1,1]}$. Thus we obtain a \emph{complete diagram of the spacetime} with accelerating NUT black holes, shown in the right part of this figure.}
\label{globalstructure}
\end{figure}

It is necessary to clarify these somewhat puzzling observations. Such an understanding of the global structure of the spacetime manifold with accelerating NUT black holes will provide us with the complete picture summarized in Fig.~\ref{globalstructure}.

To describe and investigate the \emph{complete} conformal infinity~$\scri$ of spacetimes with accelerating NUT black holes, it is much more convenient to consider the metric form (\ref{metricAC}). Similarly as for the spherical-like coordinates, it directly follows from expressions (\ref{eq:Psi}) that the corresponding curvature scalars $\Psi_A$ \emph{all vanish for} ${\,\Xi(x,y)=0}$. Such regions are thus \emph{asymptotically flat}, representing~$\scri$.
In view of the explicit form of this function (\ref{XiEXPLICIT}) it is clear that this condition is equivalent to  ${\,x-y=0\,}$. Therefore, the \emph{asymptotically flat infinity is located at}
\begin{equation}
\scri :\qquad   x=y \,, \label{loc-scri}
\end{equation}
see also Fig.~\ref{globalstructure}. The admitted range of the coordinate $x$ is ${x\in[-1,1]}$ (see the subsequent section) and thus the range of $y$ on~$\scri$ is also ${y\in[-1,1]}$. Interestingly, it is exactly the same situation as for the $C$-metric (\ref{Cmetric}), see \cite{GriffithsPodolsky:2009}.

It can now be understood, what are the specific drawbacks of the spherical-like coordinates $r,\theta$ of the metric (\ref{newmetric}) to represent~$\scri$. \emph{There is no problem in the equatorial plane} ${\theta=\frac{\pi}{2}}$ corresponding to ${x=0}$, which symmetrically divides the spacetime into two regions between the two axes (strings). Due to (\ref{loc-scri}), the scri $\scri$ in such ``transverse section'' is located at ${y=0}$, and it follows from the transformation (\ref{transformation_to_newform}) that this occurs at \emph{infinite values of}~$r$,
\begin{eqnarray}
\scri \hbox{\,\,\ at \ }\theta=\frac{\pi}{2}:\qquad   r = \pm\infty\,, \label{scri-in-r-equatoria}
\end{eqnarray}
as na\"{\i}vely assumed. However, at any other section ${\theta=\hbox{const.}}$, the conformal infinity~$\scri$ is located at \emph{finite values of}~$r$. Indeed, (\ref{loc-scri}) with (\ref{transformation_to_newform}) reads ${\cos\theta=1/[\alpha\,(r-r_{-})]}$, that is
\begin{eqnarray}
\scri \hbox{\,\,\ at any\ }\theta\ne\frac{\pi}{2}:\qquad   r = r_{-} + \frac{1}{\alpha\cos \theta }\,. \label{scri-in-r}
\end{eqnarray}
Therefore, close to the first string at ${\theta=0}$ we obtain ${r\to r_{-}+\alpha^{-1}\equiv r_a^+}$, while close to the second string at ${\theta=\pi}$ we get $r\to r_{-}-\alpha^{-1}\equiv r_a^-$, see (\ref{horizons}) and Fig.~\ref{globalstructure}.
Notice that this is exactly the condition for \emph{vanishing conformal factor} in the metric (\ref{newmetric}), (\ref{newfunctions}),
\begin{equation}
\Omega(r,\theta) = 0 \,. \label{Omega=0-scri}
\end{equation}

Such a behavior is analogous to the situation in the simpler $C$-metric \cite{GriffithsPodolsky:2009}.
However, in the present case of accelerating NUT black holes, there are \emph{two distinct asymptotically flat regions}, namely ${\scri^+}$ which is the conformal boundary of ``our universe'' in the region I$^+$, and ${\scri^-}$ which is the conformal boundary of ``parallel universe'' in the region I$^-$. In order to cover the part ${\theta>\frac{\pi}{2}}$ of ${\scri^+}$ in ``our universe'', it is necessary to also consider ${r<0}$. And vice versa: to cover the part ${\theta<\frac{\pi}{2}}$ of ${\scri^-}$ in ``parallel universe'', it is necessary to also employ ${r>0}$. This is surely possible, but quite cumbersome.

\vspace{2mm}

\subsubsection{Boost-rotation metric form and its analytic extension}
To further elucidate the global structure of the new solution (\ref{newmetric}) for accelerating NUT black holes, it is  useful to express it in a form in which its \emph{boost and rotation symmetries are explicitly manifested}. This will also provide a clear argument indicating that the analytically extended space-time represents \emph{a pair of accelerated black-hole sources}. It is achieved by applying the transformation
\begin{eqnarray}
 \zeta \rovno \frac{\sqrt{P}}{\alpha\,\Omega}\,\sqrt{|1-\alpha^2 (r - r_{-} )^2 |}\,, \label{brtrans1}\\
 \rho  \rovno \frac{\sin\theta}{\Omega}\,\sqrt{(r-r_{+})(r-r_{-})}\,, \label{brtrans2}
\end{eqnarray}
(so that ${\zeta, \rho \ge 0}$) with ${t'=\alpha\, t}$ and ${\varphi}$ unchanged. Clearly, ${\zeta=0}$ at both acceleration horizons $\HH_a^\pm$, whereas ${\rho=0}$ at both black-hole horizons $\HH_b^\pm$, and also along the two strings located at ${\theta=0}$ and ${\theta=\pi}$. An application of the transformation (\ref{brtrans1}), (\ref{brtrans2}) takes the metric (\ref{newmetric}) to the form
\begin{equation}
 \label{brdiag-metric}
  \dd s^2 = -\e^\mu\,\zeta^2\,(\dd t'- A\,\dd\varphi)^2 +\e^\lambda\,\bigl(\dd\zeta^2+\dd\rho^2\bigr)
  +\e^{-\mu}\rho^2\,\dd\varphi^2\;,
\end{equation}
where the functions ${\mu}$, ${\lambda}$ and $A$ are
\begin{eqnarray}
 \e^\mu        \rovno \frac{(r-r_{+})(r-r_{-})}{\R^2\,P}  \,,\nonumber\\[2mm]
 \e^{-\lambda} \rovno  \R^{-2} \Big( (r-r_{+})(r-r_{-})P + (m^2+l^2)\big[1-\alpha^2 (r - r_{-} )^2 \big]\sin^2\theta \Big) \,,\label{BSfunctions}\\[2mm]
 A \rovno 2\,\alpha l\, \Big(\cos\theta - \frac{\alpha}{2\sqrt{m^2+l^2}}\, \frac{r-r_{-}}{r-r_{+}} \,P\,\rho^2  \,\Big)  \,.\nonumber
\end{eqnarray}
Of course, these metric functions  need to be re-written in terms of the variables $\zeta$ and $\rho$.

When the NUT parameter vanishes, ${l=0}$, the metric becomes \emph{static} because ${A=0}$. In fact, the remaining functions $\e^\mu$ and $ \e^{-\lambda}$ then reduce exactly to expressions (14.30), (14.31) in \cite{GriffithsPodolsky:2009} for the \emph{$C$-metric}. For ${m\to0}$, the metric (\ref{brdiag-metric}) further reduces to the \emph{uniformly accelerated flat metric}, since ${\e^\mu \to 1}$ and  ${\e^{-\lambda} \to 1}$, yielding
\begin{equation} \label{accelrflat-metric}
  \dd s^2 = -\zeta^2\,\dd t'^2 + \dd\zeta^2 + \dd\rho^2 + \rho^2\,\dd\varphi^2\,.
\end{equation}
 It is equation~(14.25) in~\cite{GriffithsPodolsky:2009}, equivalent to the Bondi--Rindler metric (3.14) whose coordinates are adapted to the uniform acceleration. \emph{This weak-field limit thus provides a reasonable justification that the black-hole sources are indeed accelerating}. Moreover, in view of (\ref{brtrans1}), the acceleration is given by the parameter $\alpha$ (see also section~3.5 in~\cite{GriffithsPodolsky:2009} for more details).

Now, the metric (\ref{brdiag-metric}) in the stationary regions~II can be \emph{analytically extended through the acceleration horizons} located at ${\zeta=0}$ by transforming it to the boost-rotation symmetric form  with rotating sources (see \cite{BicakSchmidt:1989b, BicakPravda:1999, GriffithsPodolsky:2006a}). In particular, by performing the transformation\footnote{An analogous transformation in the non-stationary regions~I close to the conformal infinity~$\scri$ is ${T=\pm\zeta\cosh t'}$, ${Z=\pm\zeta\sinh t'}$.}
\begin{equation} \label{transftoTZ}
  T=\pm\zeta\sinh t'\,,\qquad  Z=\pm\zeta\cosh t' \,,
\end{equation}
the metric becomes
 \begin{eqnarray}
\dd s^2 \rovno - \frac{e^\mu}{Z^2-T^2}
  \bigg[ (Z\dd T-T\dd Z) - A\,(Z^2-T^2)\,\dd\varphi \bigg]^2  \nonumber\\[6pt]
 && \hspace{7.1mm} +e^\lambda\left[ \frac{(Z\dd Z-T\dd T)^2}{Z^2-T^2} +\dd\rho^2 \right]
      +e^{-\mu}\rho^2\, \dd\varphi^2 \,.
 \label{br-metric}
\end{eqnarray}

Clearly, ${\zeta^2 \equiv |Z^2-T^2|}$, so that the \emph{acceleration horizons  $\HH_a^\pm$ are now located at ${T=\pm Z}$}. They separate the  domains of types I and II. For the whole range of the coordinates $T$ and $Z$, the boost-rotation symmetric metric (\ref{br-metric}) covers \emph{all these regions}, with $\mu$, $\lambda$ and $A$ being specific functions of $\rho$ and ${Z^2-T^2}$, independent of $t'$ and $\varphi$.

Notice, however, that the coordinates  ${(\zeta,\rho)}$ and equivalently ${(r,\theta)}$  with the ``$+$'' sign in (\ref{transftoTZ}) each cover only \emph{half} of the section ${t'=}$~const.~corresponding to a single domain of type~II, because \emph{necessarily} ${Z>0}$. To cover also the analytically extend  regions ${Z<0}$, a \emph{second copy of these coordinates is required} by choosing the ``$-$'' sign in (\ref{transftoTZ}). This indicates that the \emph{complete spacetime actually contains a pair of uniformly accelerating NUT black holes}, similarly as in the case of the $C$-metric (see Chapter 14 in \cite{GriffithsPodolsky:2009} for the details). These two black holes accelerate away from each other, and are causally separated. The analytically extended manifold thus contains \emph{four asymptotically flat regions}, a pair of $\scri^+$ and a pair of $\scri^-$, each in ``our universe'' and in the ``parallel universe''.

Let us finally remark that at large values of the radial coordinate $r$ close to $\scri^\pm$ where ${\Omega=0}$, for any fixed value of $\theta$ the metric functions behave as ${\R \sim r}$, $P$~is a constant, and ${\Omega \sim r}$ (the case ${\theta=\frac{\pi}{2}}$ must be treated separately). It thus follows from (\ref{BSfunctions}) that the functions ${\e^\mu, \e^{-\lambda}, A}$ remain \emph{finite} in this limit, demonstrating the correct asymptotic behavior of the boost-rotation metric form (\ref{br-metric}). In fact, analogously to the procedure presented in \cite{GriffithsPodolsky:2006a}, by a properly performed rescaling of the coordinates and uniquely chosen linear combination of $t'$ and $\varphi$, for the given $\theta$  it is possible to achieve ${\e^\mu, \e^{\lambda} \to 1}$ and ${A \to 0}$ in the asymptotically flat regions of these spacetimes.

\subsection{Character of the axes ${\theta=0}$ and ${\theta=\pi}$: rotating cosmic strings}
\label{subsec:axes}

We have seen in section~\ref{subsec:horizon} that the coordinate singularities given by ${\Q(r)=0}$ represent four horizons (\ref{horizons}) associated with the Killing vector field $\partial_t$. There is also the \emph{second Killing vector field} $\partial_\varphi$, and its degenerate points identify \emph{the spatial axes of symmetry}.

They are located at the coordinate singularities of the function $\sin\theta$ in the new metric (\ref{newmetric}), and these appear at the poles ${\theta=0}$ and ${\theta=\pi}$. Therefore, the range of the spatial coordinate~$\theta$ must be constrained to ${\theta\in[0,\pi]}$. Via the simple relation ${x=-\cos\theta}$ this is equivalent to the range ${x\in[-1,1]}$ between the two poles ${x=\pm 1}$ of the function ${(1-x^2)}$ in the original form of the metric (\ref{metricAC}). The location of these poles is indicated in Fig.~\ref{globalstructure}, defining the boundary of the physical spacetime with black holes (the shaded region). Expressed in terms of the coordinates of the boost-rotation/axially symmetric metric (\ref{br-metric}), related by (\ref{brtrans2}), these poles ${\theta=0,\pi}$ correspond to ${\rho=0}$ which naturally identifies the corresponding two axes.

In analogy with the $C$-metric, such degenerate axes represent \emph{cosmic strings or struts}. Their tension is the \emph{physical source of the acceleration of the black holes}.

We have proven in Sec.~\ref{subsec:curvature} that the algebraic structure of (generic) type~I spacetime degenerates along these axes to type~D, with the only curvature component $\Psi_2$ given by (\ref{Psi2TypeD}). Subsequently, in Sec.~\ref{subsec:singularity} we have demonstrated that for ${\theta = 0}$ and ${\theta =\pi}$ the Kretschmann scalar ${{\cal K}(r) = 48 \, {\cal R}e\,(\Psi^2_2)}$ (see the expression (\ref{KretschmannEXPLIC})) is everywhere finite, as is explicitly plotted in Fig.~\ref{figureKretschmann2D} and Fig.~\ref{figureKretschmann}. There is thus \emph{no curvature singularity along these axes}. Instead, these are basically topological defects associated with \emph{conical singularities} given by \emph{deficit or excess angles} around the two distinct axes. In addition, due to the nonvanishing NUT parameter~$l$, these \emph{cosmic strings or struts are rotating}, thus introducing an internal twist to the entire spacetime with accelerating NUT black holes. We will now analyze them in more detail.

\subsubsection{Cosmic strings or struts}

We have seen that there are three explicit physical parameters of the spacetime (\ref{newmetric}), namely the mass~$m$, the acceleration~$\alpha$, and the NUT parameter $l$ of the black holes (which determine the horizon parameters ${r_\pm = m \pm \sqrt{m^2+l^2}}$, see (\ref{defr+r-}) and (\ref{horizons})). In fact, there is also the \emph{fourth free parameter} ${C}$, \emph{which is hidden in the range of the angular coordinate} ${\varphi\in[0,2\pi C)}$. It has not yet been specified. We will demonstrate its physical meaning by relating it to the deficit (or excess) angles of the cosmic strings.

Let us start with investigation of the (non)regularity of the \emph{first axis of symmetry} ${\theta=0}$ in (\ref{newmetric}). Consider a small circle around it given by ${\theta=\hbox{const.}}$, with the range ${\varphi\in[0,2\pi C)}$, assuming fixed $t$ and $r$. The invariant length of its \emph{circumference} is ${\int_0^{2\pi C}\!\! \sqrt{g_{\varphi\varphi}}\, \dd\varphi}$, while its \emph{radius} is ${\int_0^{\theta}\! \sqrt{g_{\theta\theta}}\, \dd\tilde\theta}$. The \emph{axis is regular} if their fraction in the limit ${\theta\to 0}$  is equal to ${2\pi}$. In general we obtain
\begin{equation}
 f_0 \equiv \lim_{\theta\to0} \frac{\hbox{circumference}}{\hbox{radius}}
 =\lim_{\theta\to0} \frac{2\pi C \sqrt{g_{\varphi\varphi}} }{ \theta\,\sqrt{g_{\theta\theta}}}  \,.
 \label{Accel-con0}
\end{equation}
Now, the conceptual problem is that the metric function $g_{\varphi\varphi}$ in (\ref{newmetric}), and thus the circumference, does \emph{not} approach zero in the limit ${\theta\to0}$ due to the presence of $\cos\theta$ in the first term in the metric. This problem can be resolved by the same procedure as for the classic Taub--NUT solution (see the transition between the metrics (12.1) and (12.3) in \cite{GriffithsPodolsky:2009}): By applying the transformation of the time coordinate\footnote{It leads to a closed circle instead of an open helical orbit of the axial Killing vector around ${\theta=0}$. For a recent related study of geometrical and physical properties of symmetry axes of black holes with NUT parameters see~\cite{KolarKrtous:2019}.}
\begin{equation}
t=t_{0}+2l\varphi  \,,
 \label{t-t0}
\end{equation}
the metric (\ref{newmetric}) becomes
\begin{eqnarray}
\dd s^2 \rovno  \frac{1}{\Omega^2}\,
   \Bigg[ -\frac{\Q}{\R^2} \bigg(\dd t_{0}
       + 2l \Big(2\sin^2\frac{\theta}{2} + \alpha\,\T\sin^2\!\theta \,\Big)\dd \varphi \bigg)^2 \nonumber \\
&&\hspace{23.5mm}
   + \frac{\R^2}{\Q} \,\dd r^2
   + \R^2 \bigg( \frac{\dd\theta ^2}{P} +P \sin^2\!\theta \,\dd\varphi^2 \bigg)\Bigg] , \label{newmetric-Regular-0}
\end{eqnarray}
so that
\begin{equation}
 g_{\varphi\varphi} = \frac{1}{\Omega^2}\,
   \Big[\, \R^2 P \sin^2\!\theta - 4l^2\frac{\Q}{\R^2}
   \Big(2\sin^2\frac{\theta}{2} + \alpha\,\T\sin^2\!\theta \,\Big)^2\,\Big]\,,\qquad
 g_{\theta\theta} = \frac{\R^2}{\Omega^2 P}\,   \,.
 \label{gphiphi-gthetatheta0}
\end{equation}
For very small values of $\theta$ we obtain ${ g_{\varphi\varphi} \approx \R^2 P \,\theta^2/\Omega^2}$ because the terms proportional to $l^2$ become negligible. Evaluating the limit (\ref{Accel-con0}) we thus obtain
\begin{equation}
 f_0 = 2\pi C\,\big( 1-\alpha \, (r_{+} - r_{-})\big)
     \equiv 2\pi C\,\big( 1- 2\alpha \sqrt{m^2+l^2}\,\big)  \,.
 \label{f0}
\end{equation}
\emph{The axis ${\theta=0}$ in the metric (\ref{newmetric-Regular-0}) can thus be made regular by the choice }
\begin{equation}
 C= C_0 \equiv \frac{1}{ 1- 2\alpha \sqrt{m^2+l^2}}  \,.
 \label{C0}
\end{equation}

Analogously, it is possible to regularize the \emph{second axis of symmetry} ${\theta=\pi}$. Performing the complementary transformation of the time coordinate
\begin{equation}
t=t_{\pi}-2l\varphi  \,,
 \label{t-tpi}
\end{equation}
the metric (\ref{newmetric}) becomes
\begin{eqnarray}
\dd s^2 \rovno  \frac{1}{\Omega^2}\,
   \Bigg[ -\frac{\Q}{\R^2} \bigg(\dd t_{\pi}
       - 2l \Big(2\cos^2\frac{\theta}{2} - \alpha\,\T\sin^2\!\theta \,\Big)\dd \varphi \bigg)^2 \nonumber \\
&&\hspace{23.5mm}
   + \frac{\R^2}{\Q} \,\dd r^2
   + \R^2 \bigg( \frac{\dd\theta ^2}{P} +P \sin^2\!\theta \,\dd\varphi^2 \bigg)\Bigg] , \label{newmetric-Regular-pi}
\end{eqnarray}
i.e.,
 \begin{equation}
 g_{\varphi\varphi} = \frac{1}{\Omega^2}\,
   \Big[\, \R^2 P \sin^2\!\theta - 4l^2\frac{\Q}{\R^2}
   \Big(2\cos^2\frac{\theta}{2} - \alpha\,\T\sin^2\!\theta \,\Big)^2\,\Big]\,,\qquad
 g_{\theta\theta} = \frac{\R^2}{\Omega^2 P}\,   \,.
 \label{gphiphi-gthetatheta-pi}
 \end{equation}
For ${\theta \to \pi}$ we thus obtain ${g_{\varphi\varphi} \approx \R^2 P (\pi-\theta)^2/\Omega^2}$.
The radius of a small circle around the axis ${\theta=\pi}$  is ${\int_{\theta}^{\pi}\! \sqrt{g_{\theta\theta}}\, \dd\tilde\theta}$. Evaluating the fraction
\begin{equation}
 f_\pi \equiv \lim_{\theta\to \pi} \frac{\hbox{circumference}}{\hbox{radius}}
 =\lim_{\theta\to\pi} \frac{2\pi C \sqrt{g_{\varphi\varphi}} }{ (\pi-\theta)\,\sqrt{g_{\theta\theta}}}  \,,
 \label{Accel-conpi}
\end{equation}
we obtain
\begin{equation}
 f_\pi = 2\pi C\,\big( 1+\alpha \, (r_{+} - r_{-})\big)
     \equiv 2\pi C\,\big( 1+ 2\alpha \sqrt{m^2+l^2}\,\big)  \,.
 \label{fpi}
\end{equation}
{\emph{The axis ${\theta=\pi}$ in the metric (\ref{newmetric-Regular-pi}) is thus regular for the unique choice}
\begin{equation}
 C= C_\pi \equiv \frac{1}{ 1 + 2\alpha \sqrt{m^2+l^2}}  \,.
 \label{Cpi}
\end{equation}

It is now explicitly seen that it is not possible to regularize simultaneously both the axes because ${C_0\ne C_\pi}$ and ${t_0\ne t_\pi = t_{0}+4l\varphi }$ (unless ${\alpha=0=l}$ which is just the Schwarzschild solution, regular for the standard choice ${C=1}$).

When the second axis of symmetry ${\theta=\pi}$ is made regular by the choice (\ref{Cpi}), there is necessarily a \emph{deficit angle} $\delta_0$ (conical singularity) along the first axis ${\theta=0}$, namely
\begin{equation}
 \delta_0 \equiv 2\pi-f_0 =\frac{8\pi \alpha \sqrt{m^2+l^2} }{ 1 + 2\alpha \sqrt{m^2+l^2}}>0  \,.
 \label{delta0}
\end{equation}
The corresponding tension in this \emph{cosmic string located along ${\theta=0}$ pulls the black hole, causing its uniform acceleration}. Such string \emph{extends to the full range of the radial coordinate} ${r\in (-\infty,+\infty)}$, connecting thus ``our universe'' with the ``'parallel universe'' through the nonsingular NUT black-hole interior, see Fig.~\ref{globalstructure}. Moreover, as argued in Sec.~\ref{subsection:confinity}, there is a \emph{pair} of causally separated NUT black holes accelerating away from each other by the action of two such cosmic strings, one string in each copy ${Z>0}$ and ${Z<0}$.

Complementarily, when the first axis of symmetry ${\theta=0}$ is made regular by the choice (\ref{C0}), there is necessarily an \emph{excess angle} $\delta_\pi$ along the second axis ${\theta=\pi}$, namely
\begin{equation}
 \delta_\pi \equiv 2\pi-f_\pi = -\frac{8\pi \alpha \sqrt{m^2+l^2} }{ 1 - 2\alpha \sqrt{m^2+l^2}}<0  \,.
 \label{deltapi}
\end{equation}
This represents the \emph{cosmic strut located along ${\theta=\pi}$ between the two black holes, pushing them away from each other} in opposite spatial directions ${\pm Z}$.

In particular, for black holes with vanishing NUT parameter ${l=0}$, the general results (\ref{delta0}) and (\ref{deltapi}) reduce to
\begin{equation}
 \delta_0    =  \frac{8\pi \alpha m}{ 1 + 2\alpha m} \qquad\hbox{and}\qquad
 \delta_\pi  = -\frac{8\pi \alpha m}{ 1 - 2\alpha m}  \,,
 \label{delta0pi-for-l=0}
\end{equation}
which fully agree with the known expressions for the $C$-metric, see Eqs.~(14.15)--(14.17) in \cite{GriffithsPodolsky:2009}.

\subsubsection{Rotation of these cosmic strings or struts}

With a generic NUT parameter~$l$, these \emph{cosmic strings/struts are rotating}. This can be seen by calculating the \emph{angular velocity} parameter~$\omega$ of the metric along the two different axes~\cite{ChngMannStelea:2006},
\begin{equation}
 \omega \equiv\frac{g_{t\varphi}}{g_{tt}}  \,.
 \label{omega}
\end{equation}
For the general form of the new metric (\ref{newmetric}) we obtain ${\omega =- 2l \big(\cos\theta - \alpha\,\T\sin^2\!\theta \,\big)}$. Evaluating it on the axis ${\theta=0}$ and the axis ${\theta=\pi}$, we immediately get
\begin{equation}
 \omega_0  = -2l \qquad\hbox{and}\qquad  \omega_\pi  = 2l  \,,  \label{omega0pi}
\end{equation}
respectively. Both cosmic strings/struts thus rotate. In fact, they are \emph{contra-rotating} with exactly opposite angular velocities ${\pm 2l}$ \emph{determined solely by the NUT parameter}.

If the first axis of symmetry ${\theta=0}$ is made regular by considering the metric (\ref{newmetric-Regular-0}) with the time $t_0$, then ${\omega = 2l \big(2\sin^2\frac{\theta}{2} + \alpha\,\T\sin^2\!\theta \big)}$ and the corresponding angular velocities of the axes are
\begin{equation}
 \omega_0 = 0 \qquad\hbox{and}\qquad  \omega_\pi  = 4l  \,,  \label{omega0pireg0}
\end{equation}
On the other hand, when the second axis  ${\theta=\pi}$ is regularized by switching to the metric (\ref{newmetric-Regular-pi}) with $t_\pi$, then ${\omega = - 2l \big(2\cos^2\frac{\theta}{2} - \alpha\,\T\sin^2\!\theta \big)}$ and the angular velocities of the axes are
\begin{equation}
 \omega_0 = -4l \qquad\hbox{and}\qquad  \omega_\pi  = 0  \,.  \label{omega0piregpi}
\end{equation}
Clearly, there is always a \emph{constant difference} ${\Delta\omega\equiv\omega_\pi-\omega_0=4l}$ between the angular velocities of the two rotating cosmic strings or struts, directly given by the NUT parameter~$l$.

\subsection{Regions with closed timelike curves around the rotating strings}

In the vicinity of the rotating cosmic strings or struts, which are located along ${\theta=0}$ and ${\theta=\pi}$, the spacetime with accelerating NUT black holes can serve as a specific time machine. Indeed, similarly as in the classic Taub--NUT solution, there are \emph{closed timelike curves}.

To identify these pathological causality-violating regions, let us again consider simple curves in the spacetime which are \emph{circles around the axes of symmetry} ${\theta=0}$ and ${\theta=\pi}$ such that only the periodic angular coordinate ${\varphi\in[0,2\pi C)}$ changes, while the remaining three coordinates $t$,~$r$ and~$\theta$ \emph{are kept fixed}. The corresponding tangent (velocity) vectors are thus proportional to the \emph{Killing vector field}~$\partial_\varphi$. Its norm is determined just by the metric coefficient $g_{\varphi\varphi}$, which for the general metric (\ref{newmetric}) reads
\begin{equation}
 g_{\varphi\varphi} = \frac{1}{\Omega^2}\,
   \Big[\, \R^2 P \sin^2\!\theta - 4l^2\frac{\Q}{\R^2}
   \Big(\cos\theta - \alpha\,\T\sin^2\!\theta \,\Big)^2\,\Big]   \,.
 \label{gphiphi-general}
\end{equation}
When ${l=0}$, i.e. for nonrotating cosmic strings, this metric coefficient is always positive, so that the circles are \emph{spacelike curves}. However, with the NUT parameter $l$, there are regions where ${g_{\varphi\varphi}<0}$ in which the circles (orbits of the axial symmetry) are \emph{closed timelike curves}. These pathological regions are explicitly given by the condition
\begin{equation}
\R^4 P (1-\cos^2\!\theta) <
  4l^2\Q\,\big( \cos\theta - \alpha\,\T (1-\cos^2\!\theta)\, \big)^2   \,,
 \label{gphiphi-general-condition}
\end{equation}
where the functions ${P, \Q, \T, \R}$ have been defined in (\ref{newfunctions}). Although this condition is quite difficult to be solved analytically, some general observations can easily be made.

In particular, the condition can not be satisfied in the regions where ${\Q(r)<0}$.
Assuming that the acceleration $\alpha$ is not too large, satisfying (\ref{orderofhorizonsexpl}) which implies (\ref{orderofhorizons}), the closed timelike curves can thus only appear \emph{between} the black hole horizon $\HH_b$ and the acceleration horizon $\HH_a$, that is only in the region~II$^+$ given by ${r\in(r_b^+,r_a^+)}$ or in the region~II$^-$ given by ${r\in(r_a^-,r_b^-)}$. On the contrary, the pathological domain can not occur in the region~III inside the black hole or close to the conformal infinities~$\scri^\pm$ which are the boundaries of the dynamical regions~I$^\pm$ where $r$ is temporal because ${\Q<0}$, see Fig.~\ref{globalstructure}.

These observations are nicely confirmed by  plotting the values of the relevant function $g_{\varphi\varphi}(r, \theta)$ given by (\ref{gphiphi-general}), obtained numerically for various choices of the black-hole parameters. A typical example ${m=0.5}$, ${l=3}$, ${\alpha=0.05}$ is presented in Fig.~\ref{figCTCobeosy}, for ${r>0}$ (left) and ${r<0}$ (right). The grey curves are contour lines (isolines) of a constant value of $g_{\varphi\varphi}(r, \theta)$, red color depicts large positive values, while blue color depicts negative values (dark gray domains indicate extremely large values, both positive and negative). Zeros of $g_{\varphi\varphi}$ in light yellow, determining the boundary of the pathological regions given by the condition (\ref{gphiphi-general-condition}), are exactly indicated by the thick black curves. As expected, these regions with closed timelike curves occur close to the both axes ${\theta=0}$ and ${\theta=\pi}$, were the rotating cosmic strings at located. Such regions are indeed restricted to the concentric domains (two annuli) between the black hole horizons $\HH_b^\pm$ at ${r_b^\pm = r_{\pm}}$  and the acceleration horizons $\HH_a^\pm$ at ${r_a^\pm =r_{-} \pm \alpha^{-1}}$.

\begin{figure}[h]
\centerline{
\includegraphics[height=6.5cm]{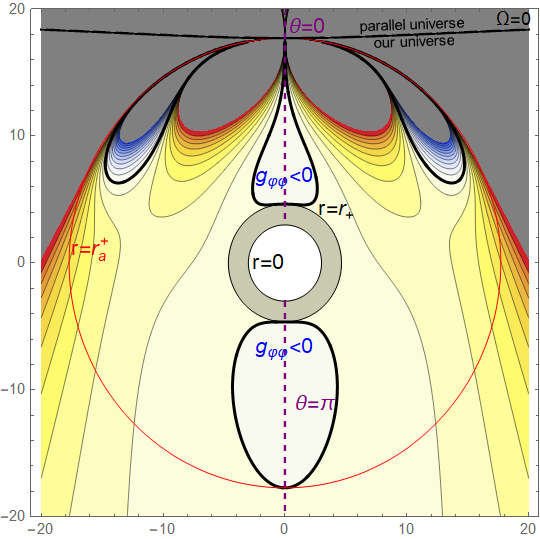}
\hspace{3mm}
\includegraphics[height=3.5cm]{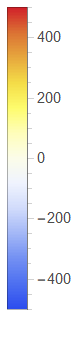}
\hspace{0mm}
\includegraphics[height=6.5cm]{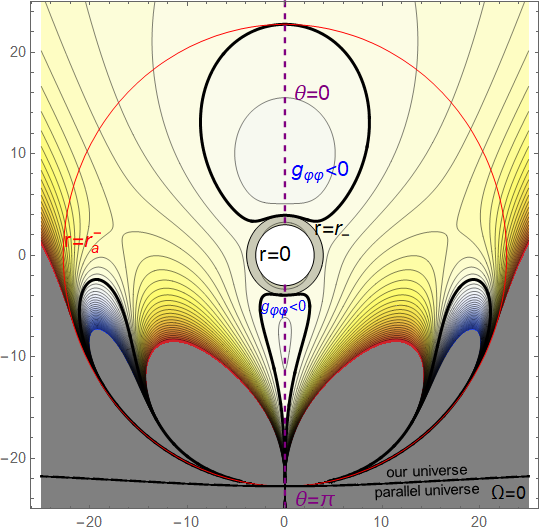}
}
\vspace{4mm}
\caption{\small
Plot of the metric function $g_{\varphi\varphi}$ (\ref{gphiphi-general}) for the general accelerating NUT black hole (\ref{newmetric}) with rotating cosmic strings on both axes ${\theta = 0}$ and ${\theta = \pi}$. Its values are visualized in quasi-polar coordinates ${{\rm x} \equiv \sqrt{r^2 + l^2}\,\sin \theta}$, ${\mathrm{y} \equiv \sqrt{r^2 + l^2}\,\cos \theta\,}$ for ${r \geq 0}$ (left) and ${r\leq 0}$ (right). The gray annulus in the center of each figure localizes the black hole bordered by its horizons $\HH_b^\pm$ at ${r_+>0}$ and ${r_-<0}$. The acceleration horizons $\HH_a^\pm$
at ${r_a^+}$ and ${r_a^-}$ (big red circles) and the conformal infinity $\scri$ at $\Omega=0$ are also shown. The grey curves are contour lines ${g_{\varphi\varphi}(r, \theta)=\hbox{const.}}$, and the values are color-coded from red (positive values) to blue (negative values). Extremely large/low values are cut and depicted in dark gray. The thick black curves in the light yellow domain are the isolines ${g_{\varphi\varphi}=0}$ determining the boundary of the pathological regions (\ref{gphiphi-general-condition}) with closed timelike curves. They occur close to both the axes ${\theta=0}$ and ${\theta=\pi}$ (purple dashed lines), but also near the acceleration horizons, forming an additional symmetric pair of ``lobes'' around ${\theta=0}$ just below~$\HH_a^+$ and around ${\theta=\pi}$ just above~$\HH_a^-$.  This plot for the choice ${m=0.5}$, ${l=3}$, ${\alpha=0.05}$ is typical.
}
\label{figCTCobeosy}
\end{figure}
\vspace{4mm}

Interestingly, for ${r>0}$ there is another pair of symmetric ``lobes'' around ${\theta=0}$ near the acceleration horizon~$\HH_a^+$ (big red circle). At a given $r$ close to ${r_a^+}$, these lobes extend to surprisingly large values of $\theta$. Similarly, there is a ``mirror'' pair of such pathological regions near $\HH_a^-$ and ${\theta=\pi}$ for ${r<0}$. In both cases, the lobes are localized around such axis, along which the acceleration horizon~$\HH_a$ closely approaches the conformal infinity~$\scri$ at~${\Omega=0}$.

In Fig.~\ref{figCTCobeosy} we visualized the regions containing the closed timelike curves for the accelerating black hole with a big value of the NUT parameter ${l=6m=3}$. However, our investigation of a large set of the parameters $m$, $l$ and $\alpha$ shows that the overall picture displayed here is quite generic.

Similarly, it is possible to investigate the regions with closed timelike curves in the special cases when one of the axes is regular. The \emph{case with regular axis} ${\theta=0}$ is described by the metric (\ref{newmetric-Regular-0}), and the corresponding metric function (\ref{gphiphi-gthetatheta0}) gives for fixed $t_0$ the condition
\begin{equation}
\R^4 P (1+\cos\theta) <
  4l^2\Q\,(1-\cos\theta)\big(\,1 + \alpha\,\T (1+\cos\theta)\, \big)^2   \,,
 \label{gphiphi-0-condition}
\end{equation}
while the complementary \emph{case with regular axis} ${\theta=\pi}$ is described by the metric (\ref{newmetric-Regular-pi}), and the corresponding metric function (\ref{gphiphi-gthetatheta-pi}) yields for fixed $t_\pi$
\begin{equation}
\R^4 P (1-\cos\theta) <
  4l^2\Q\,(1+\cos\theta)\big(\,1 - \alpha\,\T (1-\cos\theta)\, \big)^2   \,.
 \label{gphiphi-pi-condition}
\end{equation}
For a direct comparison with Fig.~\ref{figCTCobeosy}, analogous visualizations of the pathological regions in such special cases are shown in Fig.~\ref{figCTCjednaosa} for the same choice of the black-hole parameters.

\newpage
\begin{figure}[h]
\centerline{
\includegraphics[height=6.5cm]{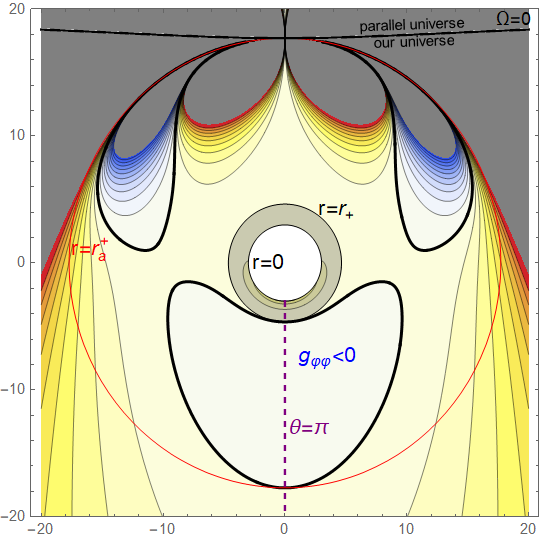}
\hspace{3mm}
\includegraphics[height=6.5cm]{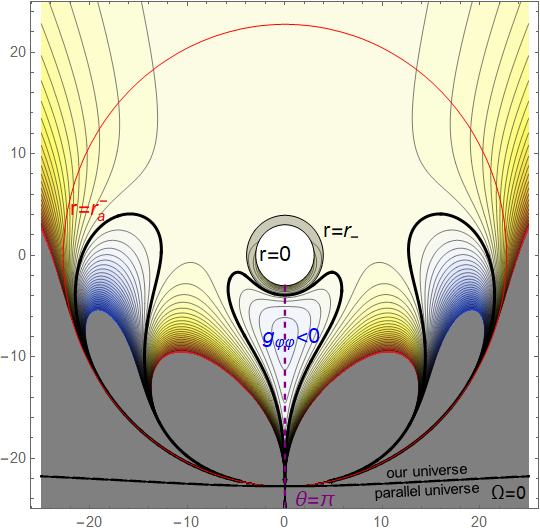}
}
\vspace{5mm}
\centerline{
\includegraphics[height=6.5cm]{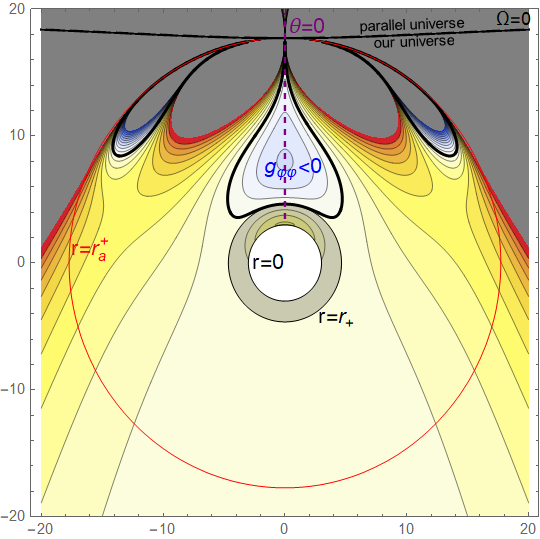}
\hspace{3mm}
\includegraphics[height=6.5cm]{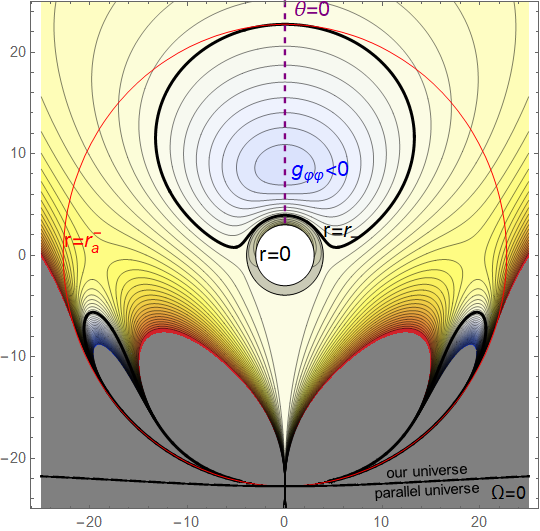}
}
\vspace{4mm}
\caption{\small
The functions $g_{\varphi \varphi}$ given by (\ref{gphiphi-gthetatheta0}) and (\ref{gphiphi-gthetatheta-pi}) for the accelerating NUT black hole metric (\ref{newmetric-Regular-0}) with the regular axis ${\theta=0}$ (top row) and for the metric (\ref{newmetric-Regular-pi}) with the regular axis ${\theta=\pi}$ (bottom row). The regions with closed timelike curves surround the remaining rotating cosmic string, and there is always an additional symmetric pair of such pathological regions near the acceleration horizons.
}
\label{figCTCjednaosa}
    \end{figure}
\vspace{3mm}

Finally, we can observe that the conditions (\ref{gphiphi-general-condition})--(\ref{gphiphi-pi-condition}) for the pathological regions simplify considerably in the \emph{absence of acceleration}. Indeed, for ${\alpha=0}$ the key functions reduce to ${P=1}$, ${\Q=\big(r-r_{+}\big)\big(r-r_{-}\big)\equiv r^2 -2m r -l^2 }$ and
${\R^2=r^2+l^2}$, see (\ref{alpha=0:newfunctions}), so that the above three conditions
(\ref{gphiphi-general-condition})--(\ref{gphiphi-pi-condition}) for the regions with closed timelike curves become, respectively,
\begin{align}
 \cos^2\theta &>  \frac{r^2+l^2}{r^2+l^2 + 4l^2f}   \,, \nonumber\\[2pt]
 \cos\theta   &< -\frac{r^2+l^2 - 4l^2f}{r^2+l^2 + 4l^2f}  \,,  \label{gphiphi-0-conditionNUT}\\[2pt]
 \cos\theta   &>  \frac{r^2+l^2 - 4l^2f}{r^2+l^2 + 4l^2f}  \,, \nonumber
\end{align}
where ${f(r) \equiv \Q/\R^2}$, see (\ref{alpha=0:finalfunction}). The result (\ref{gphiphi-0-conditionNUT}) fully agrees with the equation for the Taub--NUT spacetime presented in section 12.1.4 of the monograph~\cite{GriffithsPodolsky:2009}.

\section{Concluding summary}
\label{conclusions}

We presented and carefully investigated a remarkable class of spacetimes which represent accelerating black holes with a NUT parameter. In particular:

\begin{itemize}

\item By two independent methods we verified in Sec.~\ref{sec_vacuum} that the metric (\ref{metrikaChngMannStelea}) found by Chng, Mann and Stelea in 2006 is indeed an exact solution to Einstein's vacuum field equations.

\item To achieve this, we employed a modified version (\ref{metricAC}) of the solution in which
     one redundant parameter was removed and the original metric simplified, so that the standard $C$-metric (\ref{Cmetric}) is immediately obtained by setting the NUT-like twist parameter $\lambda$ to zero.

\item Using the metric form (\ref{metricAC}), in Sec.~\ref{sec_algebrtype} we calculated all components of the Weyl tensor in the natural null tetrad (\ref{tetrad}), namely the NP scalars $\Psi_A$ (\ref{eq:Psi}), and the corresponding curvature scalar invariants $I$ and $J$ (\ref{IJ}).

\item Since generically ${I^3 \ne 27\, J^2 }$, the Weyl tensor is of algebraically general type~I with four distinct principal null directions, explicitly given by expressions (\ref{nullrotation}) with (\ref{kvartrsol}), (\ref{kvadrsol}).

\item It explains why this class of solutions with accelerating NUT black holes has not been previously found within the large Pleba\'nski--Demia\'nski family of type~D spacetimes.

\item In Sec.~\ref{sec_newform} we derived and introduced a new metric form (\ref{newmetric}) of these solutions in ``spherical-type'' coordinates  which is much more convenient for understanding of this class of black holes.

\item In particular, its metric functions (\ref{newfunctions}), with ${r_{\pm} \equiv m \pm \sqrt{m^2+l^2}}$ given by (\ref{defr+r-}), explicitly depend on three physical parameters, namely the mass~$m$, the acceleration~$\alpha$, and the NUT parameter~$l$.

\item These black-hole parameters can be separately set to zero, recovering the well-known spacetimes in standard coordinates, namely the $C$-metric (\ref{l=0:newmetric}) when ${l=0}$, the Taub--NUT metric (\ref{alpha=0:newmetric}) when ${\alpha=0}$, the Schwarzschild metric (\ref{alpha,l=0:newmetric}), and flat Minkowski space (\ref{Minkowski}).

\item The structure of this complete family of accelerating NUT black holes is shown in Fig.~\ref{diagram}. By setting~${\alpha=0}$ or~${l=0}$, algebraically general spacetime reduces to the type~D.

\item Using the new metric (\ref{newmetric}), in Sec.~\ref{interpretation} we investigated main physical and geometrical properties of this family of accelerating NUT black holes. In particular:

\item In Subsec.~\ref{subsec:horizon} we localized the position of the horizons associated with the Killing vector field $\partial_t$. There are two black-hole horizons $\HH_b^\pm$ located at ${r_b^-\equiv r_{-}}$ and ${r_b^+ \equiv r_{+}}$ plus two acceleration horizons $\HH_a^\pm$ at ${r_a^+\equiv r_{-}+\frac{1}{\alpha}}$ and ${r_a^-\equiv r_{-}-\frac{1}{\alpha}}$. For small acceleration ${\alpha < \frac{1}{2\sqrt{m^2+l^2}}}$ they are ordered as ${r_a^-<r_b^-<0<r_b^+<r_a^+}$, see (\ref{orderofhorizons}).

\item We carefully analyzed the curvature of the spacetime in Subsec.~\ref{subsec:curvature}. We expressed the Weyl scalars (\ref{eq:Psi-spher}) in the new coordinates and frames. For
    ${l=0}$ and ${\alpha=0}$, only the Newtonian component $\Psi_2^{(r\theta)}$ remains, and its special subcases (\ref{Psi2Cmetric}) and (\ref{Psi2NUTb}) fully agree with standard expressions for the $C$-metric and the Taub--NUT metric, which are both of algebraic type~D.

\item Evaluating these Weyl scalars on the horizons, we proved that they are all regular (that is free of curvature singularities), and of a double degenerate algebraic type~D.

\item Using the curvature invariants, including the Kretschmann scalar, we proved in Subsec.~\ref{subsec:singularity} that there are no curvature singularities whenever the NUT parameter~$l$ is nonzero. This is visualized in Figs.~\ref{figureKretschmann2D} and~\ref{figureKretschmann}. Maximal (finite) values of the curvature are inside the black hole.

\item Curvature singularity appears only in the $C$-metric case ${l=0}$ at ${r=0}$. All other spacetimes in the class of accelerating NUT black holes are nonsingular, and to describe their complete manifold  it is thus necessary to consider the full range of the coordinate ${r\in(-\infty,+\infty)}$.

\item There may occur special regions in a given spacetime which are of algebraic type~D, II or~N, according to the values of the scalar curvature invariants (\ref{IJdegenregions}).

\item In Subsec.~\ref{subsection:confinity} we identified asymptotically flat regions which correspond to the conformal infinities~$\scri^\pm$ given by ${\Omega=0}$. These are simply given by the condition ${x=y}$ in the coordinates of the metric form (\ref{metricAC}).

\item Using the spherical-like coordinates of (\ref{newmetric}), the position of~$\scri^\pm$ is given by the conditions (\ref{scri-in-r-equatoria}) and (\ref{scri-in-r}), which look less intuitive.

\item All these investigations lead us to a complete understanding of the global structure of this class of spacetimes, summarized in Fig.~\ref{globalstructure}. The accelerating NUT black hole can be understood as a ``throat'' of maximal curvature which connects ``our universe'' located in the region ${r>0}$ with the second (also asymptotically flat) ``parallel universe'' in the region ${r<0}$.

\item Analytic extension across the acceleration horizons, using the boost-rotation symmetric form of the metric (\ref{br-metric}), revealed that there is actually a pair of such (causally separated) NUT black holes, which together involve four asymptotically flat regions. The two black holes uniformly accelerate in opposite directions, as in the case of the $C$-metric with ${l=0}$.

\item We clarified in Subsec.~\ref{subsec:axes} that the physical source of the acceleration of this pair of black holes is the tension (or compression) in the rotating cosmic strings (or struts) located along the corresponding two axes of axial symmetry at ${\theta=0}$ and ${\theta=\pi}$.

\item These string or struts are related to the deficit or excess angles which introduce topological defects along the axes. However, their curvature remains finite, and of algebraic type~D.

\item In general, there are string/struts along both the axes, but one of the axis can be made fully regular by a suitable choice of the constant $C$ in the range ${\varphi\in[0,2\pi C)}$. The first axis ${\theta=0}$ is regular in the metric form (\ref{newmetric-Regular-0}) with the choice (\ref{C0}), whereas the second axis ${\theta=\pi}$ is regular in the form (\ref{newmetric-Regular-pi}) with the choice (\ref{Cpi}). In the first case, there is a cosmic strut along ${\theta=\pi}$ with the excess angle (\ref{deltapi}), while in the second case there is a cosmic string along ${\theta=0}$ with the deficit angle (\ref{delta0}).

\item In addition to the deficit/excess angles, these cosmic strings/struts located along the axes of symmetry are characterized by their rotation parameter $\omega$ (angular velocity). Their values are directly related to the NUT parameter $l$, see expressions (\ref{omega0pi})--(\ref{omega0piregpi}).

\item There is always a constant difference ${\Delta\omega=4l}$ between the angular velocities of the two rotating cosmic strings or struts. If, and only if ${l=0}$, both the axes are nontwisting.

\item In the neighborhood of these rotating strings/struts there occur pathological regions with closed timelike curves. They are given by the conditions (\ref{gphiphi-general-condition})--(\ref{gphiphi-pi-condition}) and visualized in Figs.~\ref{figCTCobeosy} and~\ref{figCTCjednaosa}.

\end{itemize}

We hope that, with these geometrical and physical insights, the new explicit form (\ref{newmetric}) of the class of accelerating NUT black holes can be used as an interesting example for various types of investigations in Einstein's General Relativity, black hole thermodynamics, quantum gravity, or high-energy physics, for example by extending the recent studies \cite{AppelsGregoryKubiznak:2016, AnabalonAppelsGregoryKubiznakMannOvgun:2018}.

\section*{Acknowledgments}

This work has been supported by the Czech Science Foundation
Grant No.~GA\v{C}R 20-05421S (JP) and by the Charles University project GAUK No.~514218 (AV).

\newpage

\appendix

\newpage

\section{Curvature of general stationary axisymmetric spacetimes}
\label{appendix-A}

Let us assume a general form of stationary axisymmetric metric in coordinates ${(t, \varphi, x, y)}$ given by (\ref{axisimetric}), that is
\begin{equation}
g_{\mu \nu} =
\left( \begin{array}{cccc}
g_{tt} & g_{t \varphi} & 0 & 0 \\
g_{t \varphi}  & g_{\varphi \varphi} & 0 & 0 \\
0 & 0 & g_{xx} & 0 \\
0 & 0 & 0 & g_{yy}
\end{array} \right), \label{stationary_metric}
\end{equation}
in which all the metric functions \emph{can only depend} on $x$ and $y$. The inverse matrix is
\begin{equation}
g^{\mu \nu} =
\left( \begin{array}{cccc}
\,\,\, g_{\varphi \varphi}/D & - g_{t \varphi}/D & 0 & 0 \\
- g_{t \varphi}/D  & \quad   g_{tt}/D & 0 & 0 \\
0 & 0 & 1/g_{xx} & 0 \\
0 & 0 & 0 & 1/g_{yy}
\end{array} \right),
\end{equation}
where
\begin{eqnarray}
D \equiv g_{tt} \, g_{\varphi \varphi} - g_{t \varphi}^2\,.
\end{eqnarray}

The corresponding \emph{Christoffel symbols of the first kind} ${\Gamma_{\alpha \beta \gamma} \equiv
 {\textstyle\frac{1}{2}}\big(g_{\alpha \beta , \gamma} + g_{\gamma \alpha , \beta } - g_{\beta \gamma , \alpha}\big) }$  are
\renewcommand{\arraystretch}{1.2}
\begin{equation}
\begin{array}{llll}
\Gamma_{ttt} =0 \,,
&
\Gamma_{\varphi tt} =0 \,,
&
\Gamma_{x t t} =-{\textstyle\frac{1}{2}}g_{tt,x}\,,
&
\Gamma_{ytt} =-{\textstyle\frac{1}{2}} g_{tt,y}\,,
\\

\Gamma_{tt \varphi} =0\,,
&
\Gamma_{\varphi t \varphi} =0 \,,
&
\Gamma_{x t \varphi} =-{\textstyle\frac{1}{2}} g_{t \varphi,x}\,,
&
\Gamma_{y t \varphi} =-{\textstyle\frac{1}{2}}g_{t \varphi,y}\,,
\\

\Gamma_{ttx} ={\textstyle\frac{1}{2}}g_{tt,x}\,,
&
\Gamma_{\varphi t x} ={\textstyle\frac{1}{2}}g_{t \varphi,x} \,,
&
\Gamma_{x t x} =0 \,,
&
\Gamma_{ytx} =0\,,
\\

\Gamma_{tty} ={\textstyle\frac{1}{2}} g_{tt,y}\,,
&
\Gamma_{\varphi ty} ={\textstyle\frac{1}{2}}g_{t \varphi,y}\,,
&
\Gamma_{x ty} =0 \,,
&
\Gamma_{yty} =0 \,,
\\

\Gamma_{t \varphi \varphi} =0\,,
&
\Gamma_{\varphi \varphi \varphi} =0\,,
&
\Gamma_{x \varphi \varphi} =-{\textstyle\frac{1}{2}}g_{ \varphi \varphi,x}\,,
&
\Gamma_{y \varphi \varphi} = -{\textstyle\frac{1}{2}}g_{\varphi \varphi,y}\,,
\\

\Gamma_{t \varphi x} ={\textstyle\frac{1}{2}}g_{t \varphi,x}\,,
&
\Gamma_{\varphi \varphi x} ={\textstyle\frac{1}{2}}g_{\varphi \varphi,x}\,,
&
\Gamma_{x \varphi x} =0 \,,
&
\Gamma_{y \varphi x} =0 \,,
\\

\Gamma_{t \varphi y} ={\textstyle\frac{1}{2}}g_{t \varphi,y}\,,
&
\Gamma_{\varphi \varphi y} ={\textstyle\frac{1}{2}}g_{\varphi \varphi,y}\,,
&
\Gamma_{x \varphi y} =0 \,,
&
\Gamma_{y \varphi y} =0 \,,
\\

\Gamma_{txx} =0\,,
&
\Gamma_{\varphi x x} =0 \,,
&
\Gamma_{xxx} ={\textstyle\frac{1}{2}}g_{xx,x}\,,
&
\Gamma_{yxx} =-{\textstyle\frac{1}{2}} g_{xx,y}\,,
\\

\Gamma_{t x y} =0\,,
&
\Gamma_{\varphi xy} =0 \,,
&
\Gamma_{xxy} ={\textstyle\frac{1}{2}}g_{xx,y}\,,
&
\Gamma_{yxy} = {\textstyle\frac{1}{2}}g_{yy,x}\,,
\\

\Gamma_{t y y} =0 \,,
&
\Gamma_{\varphi yy} =0 \,,
&
\Gamma_{xyy} =-{\textstyle\frac{1}{2}}g_{yy,x}\,,
&
\Gamma_{yyy} ={\textstyle\frac{1}{2}}g_{yy,y}\,,
\\

\end{array}
\label{CHS1}
\end{equation}
\renewcommand{\arraystretch}{1}
and usual \emph{Christoffel symbols of the second kind}
${ \Gamma^\alpha_{\ \beta \gamma} \equiv g^{\alpha\sigma} \Gamma_{\sigma \beta \gamma} }$ are thus
\renewcommand{\arraystretch}{1.2}
\vspace{0.25cm}
\begin{equation}
\begin{array}{ll}
\Gamma^t_{\ tt} =0\,,			& 			\Gamma^\varphi_{\ tt} =0\,,\\

\Gamma^t_{\ t\varphi} =0\,,   	&           \Gamma^\varphi_{\ t\varphi} =0\,,\\

\Gamma^t_{\ tx} ={\textstyle\frac{1}{2}} \big( g_{\varphi \varphi} \, g_{tt,x}-g_{t \varphi} \,  g_{t \varphi,x} \big) /D  \,,
&
\Gamma^\varphi_{\ tx} ={\textstyle\frac{1}{2}}  \big( g_{tt} \,  g_{t \varphi,x} -g_{t \varphi} \,  g_{tt,x} \big)/D  \,,\\


\Gamma^t_{\ ty} ={\textstyle\frac{1}{2}}  \big(g_{\varphi \varphi} \,  g_{tt,y}-g_{t \varphi} \,  g_{t \varphi,y} \big) /D \,,
&
\Gamma^\varphi_{\ ty} ={\textstyle\frac{1}{2}} \big(g_{tt} \,  g_{t \varphi,y} -g_{t \varphi}  \, g_{tt,y} \big)/D  \,, \\


\Gamma^t_{\ \varphi \varphi} =0 \,,   & 			\Gamma^\varphi_{\ \varphi \varphi} =0 \,,\\

\Gamma^t_{\ \varphi x} ={\textstyle\frac{1}{2}} \big(g_{\varphi \varphi} \,  g_{t \varphi,x}- g_{t \varphi} \,  g_{\varphi \varphi,x} \big) /D  \,,
&
\Gamma^\varphi_{\ \varphi x} ={\textstyle\frac{1}{2}} \big(g_{tt}  \, g_{\varphi \varphi,x} -g_{t \varphi} \,  g_{t \varphi,x} \big)/D  \,,\\


\Gamma^t_{\ \varphi y} ={\textstyle\frac{1}{2}} \big(g_{\varphi \varphi} \,  g_{t \varphi,y}-g_{t \varphi} \,  g_{\varphi \varphi,y} \big) /D \,,
&
\Gamma^\varphi_{\ \varphi y} ={\textstyle\frac{1}{2}} \big(g_{tt} \,  g_{\varphi \varphi,y} -g_{t \varphi} \, g_{t \varphi,y} \big) /D  \,,\\


\Gamma^t_{\ xx} =0\,, 				& 			\Gamma^\varphi_{\ xx} =0\,,\\

\Gamma^t_{\ xy} =0\,,  				& 			\Gamma^\varphi_{\ xy} =0\,,\\

\Gamma^t_{\ yy} =0\,,  				& 			\Gamma^\varphi_{\ yy} =0\,,\\
\end{array}
\label{CHS2}
\end{equation}
\vspace{0.25cm}
\renewcommand{\arraystretch}{1}

\renewcommand{\arraystretch}{1.2}
\vspace{0.25cm}
\begin{equation}
\begin{array}{ll}
\Gamma^x_{\ tt} = -\frac{1}{2} \, g_{tt,x}/g_{xx} \,, 
&
\Gamma^y_{\ tt} = -\frac{1}{2} \, g_{tt,y}/g_{yy} 
 \,,\\

\Gamma^x_{\ t \varphi} = -\frac{1}{2} \, g_{t \varphi ,x}/g_{xx} \,, 
&
\Gamma^y_{\ t \varphi} = -\frac{1}{2} \, g_{t \varphi,y}/g_{yy} 
 \,, \\

\Gamma^x_{\ tx} =0\,, & \Gamma^y_{\ tx} =0\,,\\

\Gamma^x_{\ ty} =0\,, & \Gamma^y_{\ ty} =0\,,\\

\Gamma^x_{\ \varphi \varphi} = -\frac{1}{2} \, g_{\varphi \varphi ,x}/g_{xx} \,, 
&
\Gamma^y_{\ \varphi \varphi} = -\frac{1}{2} \, g_{\varphi \varphi,y}/g_{yy} 
 \,,\\

\Gamma^x_{\ \varphi x} =0\,, & \Gamma^y_{\ \varphi x} =0\,,\\

\Gamma^x_{\ \varphi y} =0\,, & \Gamma^y_{\ \varphi y} =0\,,\\

\Gamma^x_{\ xx} = \frac{1}{2} \, g_{xx,x}/g_{xx} \,, 
&
\Gamma^y_{\ xx} = -\frac{1}{2} \, g_{xx,y}/g_{yy} 
 \,,\\

\Gamma^x_{\ xy} = \frac{1}{2} \, g_{xx,y}/g_{xx} \,, 
&
\Gamma^y_{\ xy} = \frac{1}{2} \, g_{yy,x}/g_{yy} 
 \,,\\

\Gamma^x_{\ yy} = -\frac{1}{2} \, g_{yy,x}/g_{xx} \,, 
&
\Gamma^y_{\ yy} = \frac{1}{2} \, g_{yy,y}/g_{yy}  
 \,.\\

\end{array}
\label{CHS3}
\end{equation}
\vspace{0.25cm}
\renewcommand{\arraystretch}{1}

Now, we compute the \emph{Riemann curvature tensor}. However, instead of using the usual definition
\begin{eqnarray}
R^\mu _{\ \nu \kappa \lambda} \equiv \Gamma^\mu _{\ \nu \lambda , \kappa} - \Gamma^\mu _{\ \nu \kappa , \lambda} + \Gamma^\mu _{\ \rho \kappa} \, \Gamma^\rho _{\ \nu \lambda} - \Gamma^\mu _{\ \rho \lambda} \, \Gamma^\rho _{\ \nu \kappa} \,,
\label{Ricci_usual}
\end{eqnarray}
for our purposes we found that it is much more convenient to employ the equivalent expression
\begin{eqnarray}
R_{\mu \nu \kappa \lambda} = {\textstyle\frac{1}{2}} \big( g_{\mu \lambda, \kappa \nu} + g_{\kappa \nu , \mu \lambda} - g_{\mu \kappa , \nu \lambda} - g_{\nu \lambda , \mu \kappa} \big) + \Gamma_{\sigma \mu \lambda} \, \Gamma^\sigma _{\ \nu \kappa} - \Gamma_{\sigma \mu \kappa} \, \Gamma^\sigma _{\ \nu \lambda}  \,. \label{Riemann_formula}
\end{eqnarray}
Its advantage is that there is no need to differentiate the complicated Christoffel symbols of the second kind. This greatly simplifies subsequent computer algebra manipulations.
Direct evaluation using (\ref{CHS1}) and (\ref{CHS2}) leads to
\begin{eqnarray}
R_{t \varphi t \varphi} \rovno \frac{1}{4} \bigg( \frac{g_{t \varphi, x}^2-g_{tt,x} \, g_{\varphi \varphi ,x}}{g_{xx}} + \frac{g_{t \varphi, y}^2-g_{tt,y} \, g_{\varphi \varphi ,y}}{g_{yy}} \bigg) \,,
\nonumber \\
R_{t \varphi t x} \rovno 0 \,,
\nonumber \\
R_{t \varphi t y} \rovno 0 \,,
\nonumber \\
R_{t \varphi \varphi x} \rovno 0 \,,
\nonumber \\
R_{t \varphi \varphi y} \rovno 0 \,,
\nonumber \\
R_{t \varphi x y} \rovno \frac{1}{4\,( g_{tt} \, g_{\varphi \varphi} - g_{t \varphi}^2)}
\bigg(
g_{tt} \left( g_{t\varphi,y} \,  g_{\varphi \varphi ,x}
-g_{t\varphi,x} \,  g_{\varphi \varphi ,y}\right) \nonumber \\
&& \hspace{22mm}   -g_{t\varphi} \left(g_{tt,y} \,  g_{\varphi \varphi ,x}-g_{tt,x} \,  g_{\varphi \varphi ,y}\right)+g_{\varphi \varphi } \left(g_{tt,y} \,  g_{t\varphi,x}-g_{tt,x} \,  g_{t\varphi,y}\right)\bigg) \,, \nonumber\\
R_{txtx} \rovno - \frac{1}{2} \, g_{tt,xx}
+ \frac{1}{4} \bigg(
\frac{g_{tt} \,  g_{t \varphi ,x}^2  - 2 \, g_{t \varphi} \,  g_{tt,x} \, g_{t \varphi ,x} +  g_{\varphi \varphi} \, g_{tt,x}^2}{g_{tt} \, g_{\varphi \varphi} - g_{t \varphi}^2}  \nonumber \\
&& \hspace{22mm}  + \frac{g_{tt,x} \, g_{xx,x}}{g_{xx}}  -  \frac{g_{tt,y} \, g_{xx,y}}{g_{yy}}
\bigg) \,,
\nonumber \\
R_{txty} \rovno - \frac{1}{2} \, g_{tt,xy}
+ \frac{1}{4} \bigg(
\frac{g_{tt} \,  g_{t \varphi ,x} \,  g_{t \varphi ,y}   -  g_{t \varphi} \big( g_{tt,x} \, g_{t \varphi ,y} + g_{t \varphi ,x} \, g_{tt ,y} \big) +  g_{\varphi \varphi} \, g_{tt,x} \, g_{tt,y}}{g_{tt} \, g_{\varphi \varphi} - g_{t \varphi}^2}  \nonumber \\
&&  \hspace{22mm}  + \frac{g_{tt,x} \, g_{xx,y}}{g_{xx}}  +  \frac{g_{tt,y} \, g_{yy,x}}{g_{yy}}
\bigg) \,, \nonumber \\
R_{tx \varphi x} \rovno - \frac{1}{2} \, g_{t \varphi, xx}
+ \frac{1}{4} \bigg(
\frac{g_{tt} \,  g_{t \varphi ,x} \, g_{\varphi \varphi ,x}   - g_{t \varphi} \big( g_{t \varphi ,x}^2 + g_{tt,x} \, g_{\varphi \varphi ,x} \big)  +  g_{\varphi \varphi} \, g_{tt,x} \, g_{t \varphi ,x}}{g_{tt} \, g_{\varphi \varphi} - g_{t \varphi}^2}  \nonumber \\
&& \hspace{22mm}  + \frac{g_{t \varphi ,x} \, g_{xx,x}}{g_{xx}}  -  \frac{g_{t \varphi ,y} \, g_{xx,y}}{g_{yy}}
\bigg)\,, \label{Riemann}
\end{eqnarray}
\begin{eqnarray}
R_{tx \varphi y} \rovno - \frac{1}{2} \, g_{t \varphi, xy}
+ \frac{1}{4} \bigg(
\frac{g_{tt} \,  g_{t \varphi ,y} \, g_{\varphi \varphi ,x}   - g_{t \varphi} \big( g_{t \varphi ,x} \, g_{t \varphi ,y} + g_{tt,y} \, g_{\varphi \varphi ,x} \big)  +  g_{\varphi \varphi} \, g_{tt,y} \, g_{t \varphi ,x}}{g_{tt} \, g_{\varphi \varphi} - g_{t \varphi}^2}  \nonumber \\
&& \hspace{22mm}  + \frac{g_{t \varphi ,x} \, g_{xx,y}}{g_{xx}}  +  \frac{g_{t \varphi ,y} \, g_{yy,x}}{g_{yy}}
\bigg) \,,
\nonumber \\
R_{txxy} \rovno 0  \,,
\nonumber \\
R_{tyty} \rovno - \frac{1}{2} \, g_{tt,yy}
+ \frac{1}{4} \bigg(
\frac{g_{tt} \,  g_{t \varphi ,y}^2  - 2 \, g_{t \varphi} \,  g_{tt,y} \, g_{t \varphi ,y} +  g_{\varphi \varphi} \, g_{tt,y}^2}{g_{tt} \, g_{\varphi \varphi} - g_{t \varphi}^2}  \nonumber \\
&& \hspace{22mm}  - \frac{g_{tt,x} \, g_{yy,x}}{g_{xx}}  +  \frac{g_{tt,y} \, g_{yy,y}}{g_{yy}}
\bigg) \,,
\nonumber \\
R_{ty \varphi x} \rovno R_{tx\varphi y} - R_{t\varphi xy}\,, \nonumber \\
R_{ty \varphi y} \rovno - \frac{1}{2} \, g_{t \varphi, yy}
+ \frac{1}{4} \bigg(
\frac{g_{tt} \,  g_{t \varphi ,y} \, g_{\varphi \varphi ,y}   - g_{t \varphi} \big( g_{t \varphi ,y}^2 + g_{tt,y} \, g_{\varphi \varphi ,y} \big)  +  g_{\varphi \varphi} \, g_{tt,y} \, g_{t \varphi ,y}}{g_{tt} \, g_{\varphi \varphi} - g_{t \varphi}^2} \nonumber \\
&& \hspace{22mm}  - \frac{g_{t \varphi ,x} \, g_{yy,x}}{g_{xx}}  +  \frac{g_{t \varphi ,y} \, g_{yy,y}}{g_{yy}}
\bigg) \,,
\nonumber \\
R_{tyxy} \rovno 0 \,,
\nonumber \\
R_{\varphi x \varphi x } \rovno - \frac{1}{2} \, g_{\varphi \varphi ,xx}
+ \frac{1}{4} \bigg(
\frac{g_{tt} \,  g_{\varphi \varphi ,x}^2  - 2 \, g_{t \varphi} \,  g_{t \varphi ,x} \, g_{\varphi \varphi ,x} +  g_{\varphi \varphi} \, g_{t \varphi ,x}^2}{g_{tt} \, g_{\varphi \varphi} - g_{t \varphi}^2}  \nonumber \\
&& \hspace{22mm}  + \frac{g_{\varphi \varphi ,x} \, g_{xx,x}}{g_{xx}}  -  \frac{g_{\varphi \varphi ,y} \, g_{xx,y}}{g_{yy}}
\bigg) \,,
\nonumber \\
R_{\varphi x \varphi y} \rovno - \frac{1}{2} \, g_{\varphi \varphi ,xy}
+ \frac{1}{4} \bigg(
\frac{g_{tt} \,  g_{\varphi \varphi ,x} \,  g_{\varphi \varphi ,y}   -  g_{t \varphi} \big( g_{t \varphi ,x} \, g_{\varphi \varphi ,y} + g_{\varphi \varphi ,x} \, g_{t \varphi ,y} \big) +  g_{\varphi \varphi} \, g_{t \varphi ,x} \, g_{t \varphi ,y}}{g_{tt} \, g_{\varphi \varphi} - g_{t \varphi}^2}  \nonumber \\
&& \hspace{22mm}  + \frac{g_{\varphi \varphi ,x} \, g_{xx,y}}{g_{xx}}  +  \frac{g_{\varphi \varphi ,y} \, g_{yy,x}}{g_{yy}}
\bigg) \,, \nonumber\\
R_{\varphi xxy } \rovno 0 \,,
\nonumber \\
R_{\varphi y \varphi y } \rovno - \frac{1}{2} \, g_{\varphi \varphi ,yy}
+ \frac{1}{4} \bigg(
\frac{g_{tt} \,  g_{\varphi \varphi ,y}^2  - 2 \, g_{t \varphi} \,  g_{t \varphi ,y} \, g_{\varphi \varphi ,y} +  g_{\varphi \varphi} \, g_{t \varphi ,y}^2}{g_{tt} \, g_{\varphi \varphi} - g_{t \varphi}^2}  \nonumber \\
&& \hspace{22mm}  - \frac{g_{\varphi \varphi ,x} \, g_{yy,x}}{g_{xx}}  +  \frac{g_{\varphi \varphi ,y} \, g_{yy,y}}{g_{yy}}
\bigg) \,,
\nonumber \\
R_{\varphi y x y } \rovno 0 \,,
\nonumber \\
R_{x y x y } \rovno  - \frac{1}{2} \big(g_{xx,yy} + g_{yy,xx} \big)
+ \frac{1}{4} \bigg(
\frac{g_{xx,y}^2 + g_{xx,x} \, g_{yy,x}}{g_{xx}} + \frac{g_{yy,x}^2 + g_{xx,y} \,  g_{yy,y}}{g_{yy}} \nonumber
\bigg) \,.
\end{eqnarray}

Finally, we employ a general expression for the \emph{Ricci tensor\,},
\begin{eqnarray}
R_{\nu \lambda} \equiv g^{\mu\kappa} R_{\mu \nu \kappa \lambda} = {\textstyle\frac{1}{2}} g^{\mu\kappa}\big( g_{\mu \lambda, \kappa \nu} + g_{\kappa \nu , \mu \lambda} - g_{\mu \kappa , \nu \lambda} - g_{\nu \lambda , \mu \kappa} \big) + g^{\mu\kappa}\Gamma_{\sigma \mu \lambda} \Gamma^\sigma _{\ \nu \kappa} - g^{\mu\kappa}\Gamma_{\sigma \mu \kappa} \Gamma^\sigma _{\ \nu \lambda}, \label{Ricci_formula}
\end{eqnarray}
which yields the following non-trivial components for the Ricci tensor of the metric (\ref{stationary_metric}):
\begin{eqnarray}
R_{tt}
\rovno
- \frac{1}{2} \left(\frac{g_{tt,xx}}{g_{xx}} + \frac{g_{tt,yy}}{g_{yy}} \right)
- 2 \big( \Gamma^x _{\ t \varphi } \, \Gamma^\varphi _{\ tx} + \Gamma^y _{\ t \varphi } \, \Gamma^\varphi _{\ ty} \big)  \nonumber\\
&& - \Gamma^x _{\ tt} \big( \Gamma^t _{\ tx } - \Gamma^\varphi _{\ \varphi x } + \Gamma^x _{\ xx} - \Gamma^y _{\ xy} \big)
- \Gamma^y _{\ tt} \big( \Gamma^t _{\ ty } - \Gamma^\varphi _{\ \varphi y } - \Gamma^x _{\ xy} + \Gamma^y _{\ yy} \big),
\nonumber
\\
R_{t  \varphi}
\rovno
- \frac{1}{2} \left(\frac{g_{t \varphi ,xx}}{g_{xx}} + \frac{g_{t \varphi ,yy}}{g_{yy}} \right)
- \Gamma^x _{\ tt} \, \Gamma^t _{\ \varphi x} - \Gamma^y _{\ tt} \, \Gamma^t _{\ \varphi y}
- \Gamma^\varphi _{\ t x\ } \, \Gamma^x _{\ \varphi \varphi}- \Gamma^\varphi _{\ t y\ } \, \Gamma^y _{\ \varphi \varphi} \nonumber \\
&& - \Gamma^x _{\ t \varphi} \big( \Gamma^x _{\ xx} - \Gamma^y _{\ xy} \big) + \Gamma^y _{\ t \varphi} \big( \Gamma^x _{\ xy} - \Gamma^y _{\ yy} \big),
\nonumber
\\
R_{\varphi \varphi}
\rovno
- \frac{1}{2} \left(\frac{g_{\varphi \varphi ,xx}}{g_{xx}} + \frac{g_{ \varphi  \varphi ,yy}}{g_{yy}} \right)
- 2 \big( \Gamma^x _{\ t \varphi } \, \Gamma^t _{\  \varphi x} + \Gamma^y _{\ t \varphi } \, \Gamma^t _{\  \varphi y} \big)  \nonumber\\
&& - \Gamma^x _{\  \varphi  \varphi } \big(- \Gamma^t _{\ tx } + \Gamma^\varphi _{\ \varphi x } + \Gamma^x _{\ xx} - \Gamma^y _{\ xy} \big)
- \Gamma^y _{\  \varphi  \varphi } \big( -\Gamma^t _{\ ty } + \Gamma^\varphi _{\ \varphi y } - \Gamma^x _{\ xy} + \Gamma^y _{\ yy} \big),
\label{Ricci}
\\
R_{xx}
\rovno
- \frac{1}{2} \left( \frac{g_{\varphi \varphi} \, g_{tt,xx} -2 \, g_{t \varphi} \, g_{t \varphi ,xx} + g_{tt} \, g_{\varphi \varphi , xx}}{g_{tt} \, g_{\varphi \varphi} - g_{t \varphi}^2} + \frac{ g_{yy,xx} + g_{xx,yy}}{g_{yy}}\right)\nonumber\\
&& + \big( \Gamma^t _{\ tx } \big)^2 + 2 \, \Gamma^\varphi _{\ tx } \, \Gamma^t _{\ \varphi x } + \big( \Gamma^\varphi _{\ \varphi x } \big)^2  \nonumber\\
&& + \Gamma^x _{\ xx} \, \big( \Gamma^t _{\ tx} + \Gamma^\varphi _{\ \varphi x} \big) + \Gamma^y _{\ xy} \, \big( \Gamma^x _{\ xx}+\Gamma^y _{\ xy}\big)+\Gamma^y _{\ xx} \, \big( \Gamma^t _{\ ty}+\Gamma^\varphi _{\ \varphi y}-\Gamma^x _{\ xy}-\Gamma^y _{\ yy} \big),
\nonumber
\\
R_{xy}
\rovno
 - \frac{1}{2} \left( \frac{g_{\varphi \varphi} \, g_{tt,xy} -2 \, g_{t \varphi} \, g_{t \varphi ,xy} + g_{tt} \, g_{\varphi \varphi , xy}}{g_{tt} \, g_{\varphi \varphi} - g_{t \varphi}^2}\right) \nonumber\\
&& + \Gamma^t _{\ tx }  \Gamma^t _{\ ty } +  \Gamma^\varphi _{\ tx } \, \Gamma^t _{\ \varphi y } +  \Gamma^\varphi _{\ ty } \, \Gamma^t _{\ \varphi x } + \Gamma^\varphi _{\ \varphi x } \, \Gamma^\varphi _{\ \varphi y }  \nonumber\\
&& + \Gamma^x _{\ xy} \, \big( \Gamma^t _{\ tx} + \Gamma^\varphi _{\ \varphi x} \big) + \Gamma^y _{\ xy} \, \big( \Gamma^t _{\ ty} + \Gamma^\varphi _{\ \varphi y} \big) + \Gamma^x _{\ xy} \, \Gamma^y _{ \ xy} - \Gamma^x _{\ y y} \, \Gamma^y _{\ x x},
\nonumber
\\
R_{yy}
\rovno
- \frac{1}{2} \left( \frac{g_{\varphi \varphi} \, g_{tt,yy} -2 \, g_{t \varphi} \, g_{t \varphi ,yy} + g_{tt} \, g_{\varphi \varphi , yy}}{g_{tt} \, g_{\varphi \varphi} - g_{t \varphi}^2} + \frac{ g_{xx,yy} + g_{yy,xx}}{g_{xx}}\right) \nonumber\\
&& + \big( \Gamma^t _{\ ty } \big)^2 + 2 \, \Gamma^\varphi _{\ ty } \, \Gamma^t _{\ \varphi y } + \big( \Gamma^\varphi _{\ \varphi y } \big)^2  \nonumber\\
&& + \Gamma^y _{\ yy} \, \big( \Gamma^t _{\ ty} + \Gamma^\varphi _{\ \varphi y} \big) + \Gamma^x _{\ xy} \, \big( \Gamma^x _{\ xy}+\Gamma^y _{\ yy}\big)+\Gamma^x _{\ yy} \, \big( \Gamma^t _{\ tx}+\Gamma^\varphi _{\ \varphi x}-\Gamma^x _{\ xx}-\Gamma^y _{\ xy} \big).
\nonumber
\end{eqnarray}

\newpage

\section{Ricci tensors of conformally related metrics}
\label{appendix-B}

For the conformally related metrics (\ref{confmetric}),
\begin{eqnarray}
\tilde g_{ab}  \rovno \Omega^2\, g_{ab} \,,
\label{confmetricApB}
\end{eqnarray}
the corresponding Ricci tensors are connected as (see, e.g., \cite{Wald:book1984})
\begin{eqnarray}
 \tilde{R}_{ab} \rovno R_{ab}-2\,\Omega^{-1}\,\nabla_a\nabla_b\Omega
   -\Omega^{-1} g_{ab}\, g^{cd}\,\nabla_c\nabla_d\Omega \nonumber\\
  && \hspace{4.5mm} + 4\,\Omega^{-2}\,\nabla_a \Omega\,\nabla_b  \Omega
     -\Omega^{-2} g_{ab}\, g^{cd}\,\nabla_c \Omega \,\nabla_d \Omega  \,.
 \label{RicciRelation}
\end{eqnarray}
This implies relation between the physical and unphysical Ricci tensors $R_{ab}$ and $\tilde{R}_{ab}$, respectively,
\begin{eqnarray}
R _{ab} \rovno \,\tilde{R}_{ab} + \frac{1}{\Omega^2} \Big[ (\tilde{g}_{ab} \,\tilde{g}^{cd}
+2 \,\delta^c_a \delta^d_b) \big( \Omega_{,cd} - \tilde{\Gamma}^e_{\ cd}\,\Omega_{,e} \big) \Omega -3 \,\tilde{g}_{ab} \,\tilde{g}^{cd}\,\Omega_{,c}\,\Omega_{,d}  \Big]\,.
 \label{RicciRelationB}
\end{eqnarray}
For the metric (\ref{confmetric}), (\ref{conformalmetric}), the conformal factor (\ref{Omegaaxisym}) is independent of $\varphi$ and $t$, so that the resulting metric is again stationary and axisymmetric, in which case the relations (\ref{RicciRelationB}) simplify to
\begin{eqnarray}
R_{tt} \rovno \,\tilde{R}_{tt} + \frac{\Phi}{\Omega} \, \tilde{g}_{tt} - \frac{2}{\Omega} \big( \tilde{\Gamma}^x _{\ tt} \, \Omega_{, x} + \tilde{\Gamma}^y _{\ tt} \, \Omega_{,y} \big)\,,
\nonumber \\
R_{t \varphi} \rovno \,\tilde{R}_{t \varphi} + \frac{\Phi}{\Omega} \, \tilde{g}_{t \varphi} - \frac{2}{\Omega} \big( \tilde{\Gamma}^x _{\ t \varphi} \, \Omega_{, x} + \tilde{\Gamma}^y _{\ t \varphi} \, \Omega_{,y} \big)\,,
\nonumber \\
R_{\varphi \varphi} \rovno \tilde{R}_{\varphi \varphi} + \frac{\Phi}{\Omega} \, \tilde{g}_{\varphi \varphi} - \frac{2}{\Omega} \big( \tilde{\Gamma}^x _{\ \varphi \varphi} \, \Omega_{, x} + \tilde{\Gamma}^y _{\ \varphi \varphi} \, \Omega_{,y} \big)\,,  \nonumber \\
R_{xx} \rovno \,\tilde{R}_{xx} + \frac{\Phi}{\Omega} \, \tilde{g}_{xx} + \frac{2}{\Omega} \big( \Omega_{,xx}-\tilde{\Gamma}^x _{\ xx} \, \Omega_{, x} - \tilde{\Gamma}^y _{\ xx} \, \Omega_{,y} \big)\,,
\label{Ricci_conformal} \\
R_{xy} \rovno \,\tilde{R}_{xy} + \frac{\Phi}{\Omega} \, \tilde{g}_{xy} + \frac{2}{\Omega} \big( \Omega_{,xy}-\tilde{\Gamma}^x _{\ xy} \, \Omega_{, x} - \tilde{\Gamma}^y _{\ xy} \, \Omega_{,y} \big)\,,
\nonumber \\
R_{yy} \rovno \,\tilde{R}_{yy} + \frac{\Phi}{\Omega} \, \tilde{g}_{yy} + \frac{2}{\Omega} \big( \Omega_{,yy}-\tilde{\Gamma}^x _{\ yy} \, \Omega_{, x} - \tilde{\Gamma}^y _{\ yy} \, \Omega_{,y} \big) \,,
\nonumber
\end{eqnarray}
where
\begin{eqnarray}
\Phi \equi - \frac{1}{\tilde{D}} \Big[ \big(   \tilde{g}_{\varphi \varphi} \tilde{\Gamma}^x_{\ tt} - 2 \, \tilde{g}_{t \varphi}  \tilde{\Gamma}^x_{\ t\varphi}  + \tilde{g}_{tt}  \tilde{\Gamma}^x_{\ \varphi \varphi} \big)  \, \Omega_{,x}  + \big(   \tilde{g}_{\varphi \varphi} \tilde{\Gamma}^y_{\ tt} - 2 \, \tilde{g}_{t \varphi}  \tilde{\Gamma}^y_{\ t\varphi}  + \tilde{g}_{tt}  \tilde{\Gamma}^y_{\ \varphi \varphi} \big)  \,  \Omega_{,y}  \Big] \label{Phi}\\
&& \!\!\!\! + \frac{1}{\tilde{g}_{xx}} \big( \Omega_{xx}- \tilde{\Gamma}^x _{\ xx} \Omega_{,x} - \tilde{\Gamma}^y _{, xx} \Omega_{,y} \big) + \frac{1}{\tilde{g}_{yy}} \big( \Omega_{yy}- \tilde{\Gamma}^x _{, yy} \Omega_{,x} - \tilde{\Gamma}^y _{\ yy} \Omega_{,y} \big)
 - \frac{3}{\Omega} \bigg( \frac{\Omega_{,x} ^2}{\tilde{g}_{xx}} + \frac{\Omega_{,y} ^2}{\tilde{g}_{yy}} \bigg) \,, \nonumber
\end{eqnarray}
and the determinant for the metric (\ref{conformalmetric}) reads
\begin{eqnarray}
\tilde{D} \equi \tilde{g}_{tt} \, \tilde{g}_{\varphi \varphi} - \tilde{g}_{t \varphi}^2 = \Omega^4 D \nonumber\\
\rovno - (x-y)^8 \, (1-x^2)^3 F^3(x) \, (y^2-1)^3 F^3(y) \,  \tilde{H}^2(x,y) \,.
\label{deftildeD}
\end{eqnarray}

\end{document}